\documentclass[10pt,twocolumn]{article}
\usepackage[a4paper,margin=0.7in,columnsep=0.28in]{geometry}
\usepackage{amsmath,amssymb,amsthm,graphicx,bm,booktabs,array}
\graphicspath{{figures/}}
\usepackage{tikz}
\usetikzlibrary{quantikz2}
\usepackage[colorlinks=true,linkcolor=blue,citecolor=blue,urlcolor=blue]{hyperref}
\usepackage{setspace}
\usepackage{caption}
\usepackage{microtype}
\newcommand{\Pf}{\mathrm{Pf}}
\providecommand{\ket}[1]{|#1\rangle}
\providecommand{\bra}[1]{\langle#1|}
\newcommand{\avg}[1]{\langle#1\rangle}
\newcommand{\Gam}{\Gamma}
\newcommand{\NC}{\mathrm{NC}^0}
\newtheorem{theorem}{Theorem}
\newtheorem{lemma}{Lemma}
\newtheorem{conjecture}{Conjecture}

\begin{document}
\twocolumn[{%
\begin{center}
{\LARGE\bfseries Genuine multipartite nonlocality is fermionic magic}\\[1.2em]
{\large Jamal Slim}\\[0.3em]
Deutsches Elektronen-Synchrotron DESY, 22603 Hamburg, Germany\\
\texttt{jamal.slim@desy.de}
\end{center}

\vspace{1em}
\noindent\textbf{Free fermion dynamics is classically simulable, and a single non-Gaussian state makes it universal. We show that the same boundary governs quantum nonlocality. Measured as fermionic qubits, fermionic Gaussian states reach at most the biseparable bound of the Mermin inequality for three parties and, for any number of parties, violate the Mermin family by exactly the factor of a single entangled pair. They are never genuinely multipartite nonlocal. Genuine multipartite nonlocality and computational universality are thus one resource for fermions, the opposite of the stabilizer case. The mechanism is fermion parity. The bound is monogamous and certifies non-Gaussianity from measured correlations alone. On superconducting processors we watch nonlocality beyond free fermions switch on under an interaction between densities, confirm the monogamy law on 96 qubits, and measure the ground state of twelve interacting fermions on a ring cross the bound as the interaction grows, detecting the charge density wave transition.}

\vspace{1em}
\vspace{1.2em}
}]

\section*{Main}

Two families of quantum dynamics can be simulated efficiently on a classical computer, the Clifford group
and free fermions, and in both a single additional resource restores universality. For Clifford circuits
the resource is magic, and a sharp question is what it buys physically. Nonlocality is not the answer,
since Mermin's GHZ experiment~\cite{Mermin1990} is itself a Clifford experiment and reaches the algebraic
maximum of genuine multipartite nonlocality while remaining classically simulable. For stabilizer
computation maximal nonlocality is free, and in odd dimensions the resource separating it from universal
computation has been identified instead with contextuality~\cite{Howard2014}.

Free fermions are the other simulable family. Matchgate circuits, the nearest neighbour gates generated by
quadratic Majorana Hamiltonians, are efficiently
simulable~\cite{Valiant2002,TerhalDiVincenzo2002,Knill2001,BravyiKitaev2002,JozsaMiyake2008}, are equivalent
to space bounded quantum computation~\cite{JozsaKrausMiyakeWatrous2010}, and sit one gate away from
universality~\cite{BrodChilds2014,Beenakker2004charge}. Every pure non-Gaussian fermionic state promotes them
to universality~\cite{Hebenstreit2019}, so fermionic non-Gaussianity, fermionic magic, is exactly the resource
separating the two regimes, and a resource theory has grown around
it~\cite{DiasKoenig2024,ReardonSmith2024,CudbyStrelchuk2023,Hakkaku2022,Sierant2026,Haug2026}. Two notions
share this vocabulary. Recent works study the nonstabilizerness of fermionic states and its nonlocal part~\cite{Malvimat2026,Iannotti2026}. Here magic means non-Gaussianity, and nonlocality means Bell
nonlocality. Free fermions are known to be nonlocal. Mesoscopic Bell tests were proposed two decades
ago~\cite{Samuelsson2003,Beenakker2004}, and matchgate settings reach the Tsirelson value of CHSH between two
parties~\cite{Clarke2016}. The topologically protected subset, braiding and fusion of Majoranas, cannot violate
CHSH on two encoded qubits~\cite{HowardVala2012} and cannot reach the GHZ paradox of three parties~\cite{CampbellHobanEisert2014}. How much genuinely multipartite nonlocality a fermionic Gaussian state
can carry, over the full continuous manifold, has had no answer.

We show that the fermionic case is the opposite of the stabilizer one. Under local bilinear readout whose two
settings share a direction, the case realized by any dual rail or Majorana qubit, the Mermin correlator of three parties for any fermionic Gaussian state is at most $2\sqrt2$, exactly the biseparable bound, and the bound
is attained. Free fermions reach the ceiling available to two entangled parties with a third acting alone,
and stop there. The law holds for every number of parties, the maximum being $2^{n/2}$, so free fermions carry
the Bell violation of a single Tsirelson pair, $\sqrt2$, however many parties share the state, against
$2^{(n-1)/2}$ allowed by quantum mechanics. Where stabilizer states attain maximal genuine multipartite
nonlocality for free, fermionic Gaussian states cannot attain it at all. Read together with
ref.~\cite{Hebenstreit2019}, this identifies genuine multipartite nonlocality and computational universality
as the same resource within fermionic linear optics. The bound is monogamous, which separates two families of
constant depth circuits differing in the angle of one gate exponentially, and it certifies non-Gaussianity
from measured outcomes with no model of the state. We test all of this on four IBM processors of two
architectures, and we put the bound to a physical model, where it detects the charge density wave transition
of interacting fermions on a ring with a robustness that grows exponentially with the ring, and measure that
too.

\subsection*{One pair of Bell nonlocality}

Party $j$ holds a block of Majorana modes and for each of two settings measures a Majorana bilinear with outcomes $\pm1$ $O=i(u\!\cdot\!c)(v\!\cdot\!c)$, the occupation of one mode after a local Gaussian rotation, which is
the natural readout of fermionic linear optics (Methods). Its two settings share a direction when they have
the form $O_j^{0}=i(u_j\!\cdot\!c)(v_j^{0}\!\cdot\!c)$ and $O_j^{1}=i(u_j\!\cdot\!c)(v_j^{1}\!\cdot\!c)$, and
$\varphi_j$ is the angle between $v_j^0$ and $v_j^1$. With $Q_j=O_j^0+iO_j^1$ the Mermin polynomial is
$M_n=\mathrm{Re}\prod_jQ_j$, with local bound $2^{\lfloor n/2\rfloor}$ and algebraic maximum
$2^{n-1}$~\cite{Mermin1990}. For three parties the largest value of biseparable quantum states is
$2\sqrt2$~\cite{Collins2002,Bancal2011}, and Svetlichny's polynomial $S_3=\sqrt2\,\mathrm{Re}(e^{-i\pi/4}Q_AQ_BQ_C)$
is bounded by $4$ in every model in which one party is local and reaches $4\sqrt2$ quantum
mechanically~\cite{Svetlichny1987,SeevinckSvetlichny2002}.

\textbf{Theorem 1.}\ \emph{For every fermionic Gaussian state on any number of modes, and all local Majorana
bilinear settings in which each party's two measurement planes either share a direction or commute,}
\begin{equation}
  \bigl|\avg{Q_AQ_BQ_C}\bigr|\ \le\ 2\sqrt{2},
  \label{eqmain}
\end{equation}
\emph{and the bound is attained. If all three parties use anticommuting settings the bound tightens to $2$.}

\textbf{Corollary.}\ \emph{Under the same hypothesis, $|\avg{M_3}|\le2\sqrt2$ and $|\avg{S_3}|\le4$. No fermionic
Gaussian state can certify genuine tripartite entanglement independently of the device through the Mermin
inequality, and none violates the Svetlichny inequality.}

The corollary is immediate, since $M_3$ and $S_3/\sqrt2$ are real parts of $\avg{Q_AQ_BQ_C}$ times
unit modulus phases. The mechanism is parity. Reaching the algebraic maximum requires coherence between
``all empty'' and ``all full'' on three auxiliary modes, and for any Gaussian state on an odd number of modes
the product of those two probabilities is at most $2^{-M-1}$ rather than $2^{-M}$, because the relevant
spectrum is doubly degenerate and must contain a real eigenvalue (Methods). Gaussian states are built from
pairs, and three is odd.

\emph{Encoded qubits.} The hypothesis is automatic for qubits encoded in fermions. In a dual rail or Majorana tetron qubit the
logical Paulis are bilinears in three Majoranas, for dual rail $Z_L=-ic_1c_2$, $X_L=-ic_2c_3$ and $Y_L=ic_1c_3$, so
every logical measurement $\bm n\cdot\bm\sigma_L$ is a single bilinear whose plane lies in a three dimensional space,
and two planes in three dimensions always share a direction. For such parties Theorems~1 to~4 hold for every choice
of local measurements, and $\varphi_p$ is simply the angle between party $p$'s two Bloch vectors (Methods). Only parties
that measure general bilinears over four or more Majoranas fall outside the hypothesis.

Theorem~1 states the maximum. The proof determines the value itself, at every setting, and it is attained
everywhere. Bring party $p$'s settings to the canonical form $O_p^0=ic_1c_2$ and
$O_p^1=ic_1(\cos\varphi_p\,c_2+\sin\varphi_p\,c_3)$ by a local rotation.

\textbf{Theorem 2.}\ \emph{For settings sharing a direction with angles $\varphi_p\in[0,\pi/2]$, over all
fermionic Gaussian states on any number of modes,}
\begin{multline}
\max\bigl|\avg{Q_AQ_BQ_C}\bigr|=B(\varphi)\\
\equiv2\sqrt2\,\max_{p}\,\cos\tfrac{\varphi_p}{2}\cos\tfrac{\varphi_q-\varphi_r}{2},
\label{eqBphi}
\end{multline}
\emph{where $\{q,r\}$ are the two parties other than $p$. The maximum is attained by an explicit pure
product state, and $B=2\sqrt2$ if and only if some party has $\varphi_p=0$ and the other two have equal
angles.}

The free fermion value is largest not at one point but on a family, one party passive and the other two at
any common angle, and it degrades smoothly as the angles separate (Methods). Its minimum over all angles is
the local bound $2$, reached at four settings only. The constant is one instance of a law for every $n$, with
a proof independent of the argument for three parties.

\textbf{Theorem 3.}\ \emph{Let $n$ parties hold disjoint sets of modes, and let each party $j$ measure two
bilinears sharing a direction $u_j$, $O_j^0=i(u_j\!\cdot c)(v_j^0\!\cdot c)$ and $O_j^1=i(u_j\!\cdot c)
(v_j^1\!\cdot c)$. Then for every fermionic Gaussian state, pure or mixed, on any number of modes,}
\begin{equation}
\Bigl|\Bigl\langle\prod_{j=1}^{n}Q_j\Bigr\rangle\Bigr|\le2^{n/2},
\label{eqnlaw}
\end{equation}
\emph{attained by a product of the two party states of Theorem~2, with one lone party at angle zero if $n$ is
odd.}

The proof is five lines (Methods). Each party contributes the norm of one complex form, $\|\ell_j\|^2=2$ with
$\ell_j=v_j^0+iv_j^1$, and a Gaussian state cannot amplify it because its covariance is a contraction. The
Mermin polynomial and its rotated form $\sqrt2\,\mathrm{Re}(e^{-i\pi/4}\prod_jQ_j)$, which for $n=2$ is CHSH,
have local bounds $2^{\lfloor n/2\rfloor}$ and $2^{\lceil n/2\rceil}$. Theorem~3 bounds them by $2^{n/2}$ and
$2^{(n+1)/2}$, both attained. Free fermions therefore violate whichever of the two is the stronger
inequality by exactly the factor $\sqrt2$ and never the other, however many parties share the state, while
quantum mechanics allows $2^{(n-1)/2}$. The Svetlichny polynomial of $n$ parties has hybrid local bound
$2^{n-1}$~\cite{SeevinckSvetlichny2002}, and Theorem~3 gives $|S_n|\le\sqrt2\cdot2^{n/2}\le2^{n-1}$ for every
$n\ge3$. Free fermions are never genuinely multipartite nonlocal for any number of parties, with equality only
at $n=3$. For even $n$ the free fermion bound is the local bound itself, so at four or six parties any
violation of the Mermin inequality certifies magic, and we measure exactly that below.

\subsection*{The resource of universality}

The constant of Theorem~1 is the quantum bound for biseparable states of the Mermin inequality, and a value above it
witnesses genuine tripartite entanglement independently of the device~\cite{Collins2002,Bancal2011}. It is not a bound
on genuine multipartite nonlocality, since a Popescu-Rohrlich box shared by two parties plus a deterministic
third reaches $M_3=4$~\cite{Bancal2011}. That is why Theorem~1 is stated for the modulus, which controls every
phase rotated Mermin polynomial at once, and $|\avg{Q_AQ_BQ_C}|\le2\sqrt2$ is exactly Svetlichny's bound
$S_3\le4$. Free fermions saturate that boundary and never cross it. The attaining states of Theorem~2 make
the reason concrete. They are biseparable, two parties at the Tsirelson point of a Bell expression with two settings and the third acting alone. Combined with ref.~\cite{Hebenstreit2019}, which shows that any pure
non-Gaussian state promotes matchgates to universality and hence to the algebraic maximum, this gives a
two sided statement. Within fermionic linear optics, a certificate of genuine tripartite
nonlocality is a certificate of fermionic magic, and every fermionic magic state enables one.

\subsection*{Monogamy and shallow circuits}

A single cell gives a constant gap. The bound multiplies.

\textbf{Theorem 4.}\ \emph{Let $m$ Mermin cells occupy pairwise disjoint sets of modes, let $\rho$ be one
arbitrary global Gaussian state on all of them, arbitrarily correlated across cells, and let each party
measure a bilinear fixed by its setting, the settings of each cell drawn independently and uniformly from
its four even weight inputs. Let $W_i$ be the event that cell $i$ wins. If every cell satisfies the
hypothesis of Theorem~1, then}
\begin{equation}
  \Pr\Bigl[\,\textstyle\bigcap_{i=1}^{m}W_i\Bigr]\le\beta^{m},\qquad \beta=\frac{2+\sqrt2}{4} .
\end{equation}
\emph{For arbitrary bilinear settings the same holds with $\beta'=\tfrac12+\tfrac{3.26}{8}=0.9075$, a
computer certified constant, and numerically with $\tfrac12+\tfrac{3.12}{8}$.}

The proof (Methods) rests on the fact that conditioning a Gaussian state on occupation outcomes leaves a
Gaussian state, so a Gaussian strategy cannot bank correlations in one cell to pay for another. Direct
maximization over one global mixed Gaussian state on two cells, with unrestricted covariance between the cells,
converges to $\beta^2$ from random initializations and never exceeds it. Theorem~4 converts into a circuit
statement. For a constant depth circuit of nearest neighbour matchgates acting on an input in the computational basis,
with the parties' input bits loaded at qubits with disjoint light cones, each party sees one fixed Gaussian
state, an orthogonal transformation in its cone that depends on the setting and occupation readout, so Theorem~4 applies
(Methods). There is therefore a relational problem, $m$ Mermin cells at pairwise distance exceeding twice the
depth, that a constant depth circuit of matchgates plus controlled phase gates solves with probability one,
while every constant depth matchgate circuit succeeds with probability at most $\beta^{m}$ under the hypothesis
of Theorem~1, and every $\NC$ circuit at most $0.75^m$~\cite{Mermin1990,BGK2018}. Both families are
geometrically local, constant depth and classically simulable, and they differ in the phase of one gate.
This is the free fermion counterpart of the Clifford plus magic theorem on shallow circuits of ref.~\cite{Zhang2024},
with an exponentially rather than inverse polynomially small success probability.

The bound is also a witness of a different kind from the state based measures of non-Gaussianity, the
extent, the robustness and the fermionic nonlinearity~\cite{DiasKoenig2024,ReardonSmith2024,Hakkaku2022}. If
three parties performing local fermionic linear optical readout observe a Mermin value above $2\sqrt2$, no
fermionic Gaussian state, pure or mixed, on any number of modes, can have produced it. The certificate uses
only the table of outcomes, with no model of the state and no tomography, and it holds with no assumption on
the noise through Hoeffding's inequality (Methods). Noise rescales the ideal signal by a visibility $A$, and
certification at the operating point requires $A>1/\sqrt2$ for three parties, $A>1/2$ for four and $A>1/4$
for six, a threshold that falls with the number of parties because the bound grows only as the square root
of the signal.

\subsection*{Experiments}

Encode each logical qubit in two adjacent physical qubits, a dual rail with code space $\{\ket{01},\ket{10}\}$,
so that $Z_L=Z_{r_0}$, $X_L=X_{r_0}X_{r_1}$ and $Y_L=Y_{r_0}X_{r_1}$ are Majorana bilinears and every logical
Clifford and Pauli readout is Gaussian (Fig.~\ref{figcircuit}). The single non-Gaussian element is the
controlled phase between dominoes $\mathrm{CP}(\varphi)$, a matchgate only at $\varphi=0$, which in fermionic terms is
$e^{i\varphi\,n_rn_{r'}}$, an interaction between the densities of two modes applied for a time $\varphi$ in units
of its strength. The dial is the interaction. With the settings of Methods the ideal signal is exactly
$M(\varphi)=2(1-\cos\varphi)$, so the free fermion bound is crossed at $\varphi^*=2\arcsin2^{-1/4}=114.47^\circ$,
beyond which the data cannot be reproduced by any free fermion circuit with local bilinear readout. Every
logical measurement of a dual rail qubit is a bilinear in three Majoranas, so the compiled settings satisfy
the hypothesis of Theorem~1 exactly, and all values reported are raw, with no postselection, readout
mitigation or extrapolation (Methods).

\begin{figure}[t]
	\begin{quantikz}[column sep=0.44cm, row sep={0.60cm,between origins}]
		\lstick[2]{$A$} & \qw      & \gate[2]{H_L} & \qw
		& \gate[2]{B_A} & \meter{} \\
		& \gate{X} &               & \ctrl{1}
		&               & \meter{} \\
		\lstick[2]{$B$} & \qw      & \gate[2]{H_L} & \gate[style={fill=red!14}]{P(\varphi)}
		& \gate[2]{B_B} & \meter{} \\
		& \gate{X} &               & \ctrl{1}
		&               & \meter{} \\
		\lstick[2]{$C$} & \qw      & \gate[2]{H_L} & \gate[style={fill=red!14}]{P(\varphi)}
		& \gate[2]{B_C} & \meter{} \\
		& \gate{X} &               & \qw
		&               & \meter{}
	\end{quantikz}
	\caption{\textbf{The cell.} Three dual rail logical qubits $A,B,C$ on six physical qubits, the upper wire of each
		pair being $r_0$. $X$ prepares $|0\rangle_L=|01\rangle$. $H_L=R_z(\tfrac\pi2)R_{XX}(\tfrac\pi2)R_z(\tfrac\pi2)$
		and the basis rotations $B_j\in\{\mathbb{I},H_L,S_L^{\dagger}H_L\}$ are matchgates. The only non-Gaussian
		elements are the two controlled phases $P(\varphi)$ (shaded) between neighboring dominoes, and at
		$\varphi=0$ the whole circuit is a matchgate circuit.}
	\label{figcircuit}
\end{figure}

\begin{figure}[t]
	\centering
	\includegraphics[width=\linewidth]{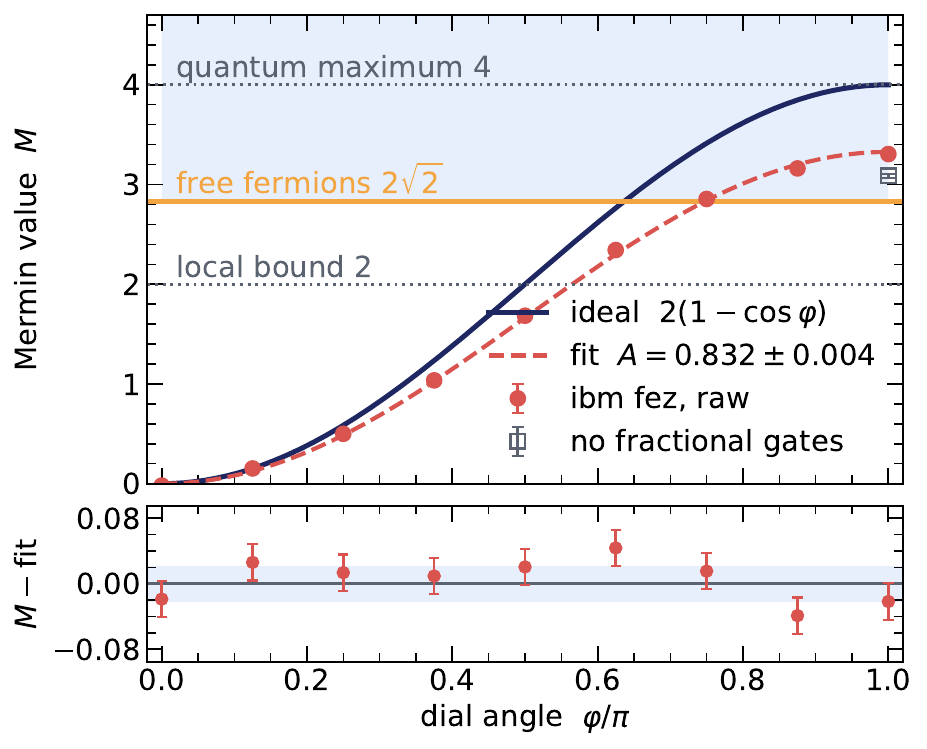}
	\caption{\textbf{The dial.} One cell on \texttt{ibm\_fez}, $8192$ shots per setting, raw data with no postselection. $M$
		against $\varphi$, with the local bound $2$, the Gaussian maximum $2\sqrt2$ (Theorem~1) and the quantum
		maximum $4$. The solid curve is the ideal $2(1-\cos\varphi)$ and the dashed curve the fit with one parameter
		$A\,2(1-\cos\varphi)$ with $A=0.832\pm0.004$. The open square is a repeat compiled without fractional
		gates. The lower panel shows the fit residuals against the $\pm\sigma(M)$ band. They sit at the
		shot noise level, so the loss is a single visibility independent of $\varphi$ and not a distortion of the
		cosine. Error bars are $\sigma(M)=2/\sqrt N$ at worst case variance and are smaller than the markers.}
	\label{figdial}
\end{figure}

One cell on \texttt{ibm\_fez}, swept over nine values of $\varphi$ with $8192$ shots per setting, gives a raw
Mermin value rising monotonically to $3.305\pm0.022$ at $\varphi=\pi$, $0.48$ above $2\sqrt2$
(Fig.~\ref{figdial}). A fit with one parameter $M=A\,2(1-\cos\varphi)$ returns a visibility $A=0.832\pm0.004$ independent of $\varphi$, with residuals at the shot noise level, so the measured curve is the predicted one
rescaled. Three independent jobs at $\varphi=\pi$ gave $3.340$, $3.243$ and $3.305$, all certifying. The same
scan on \texttt{ibm\_berlin}, a different architecture, gives $3.538\pm0.022$ with $A=0.869\pm0.005$. Both values
certify under the bound that assumes nothing about the noise at confidence $1-10^{-9}$.

\begin{figure}[t]
	\centering
	\includegraphics[width=\linewidth]{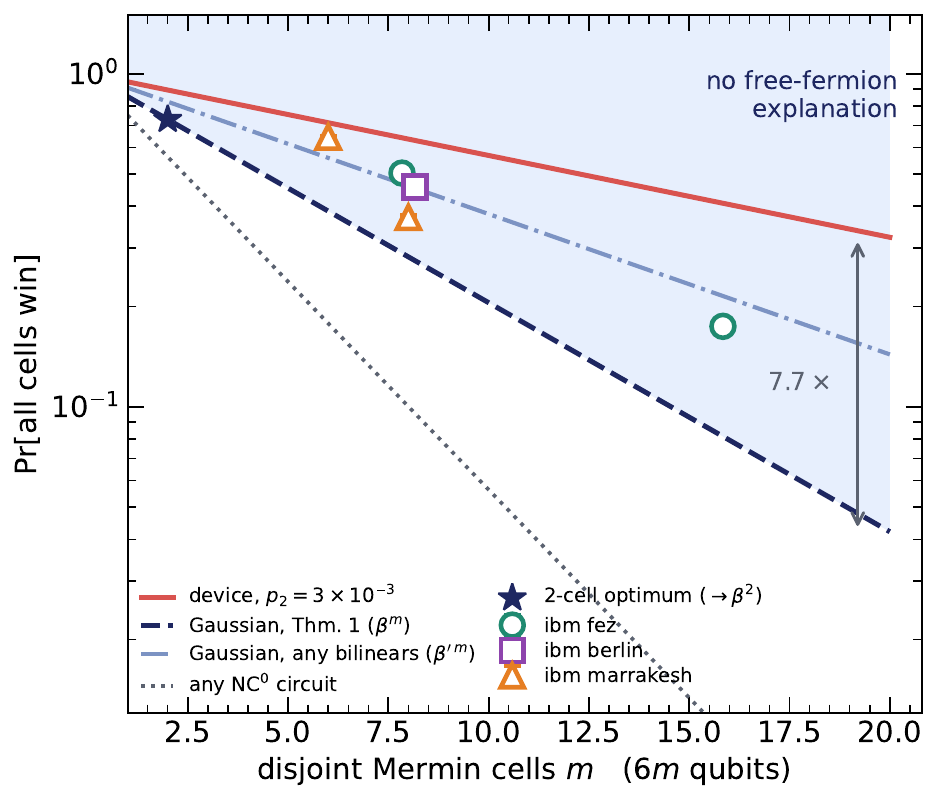}
	\caption{\textbf{Monogamy.} Joint success probability over $m$ disjoint Mermin cells, $6m$ qubits. The curves are
		theory, namely the Gaussian bound $\beta^m$ of Theorem~4 under the hypothesis of Theorem~1 (dashed),
		$\beta'^{\,m}$ for arbitrary bilinear settings with the certified constant $3.26$ (dashed and dotted), the $\NC$
		bound $0.75^m$ (dotted), and the $\varphi=\pi$ circuit at Heron class noise, $p_2=3\times10^{-3}$ and
		$1.5\%$ readout, $0.945^m$ (solid), a factor $7.7$ above $\beta^m$ at $m=20$. The star is the optimum for two cells, the
		Gaussian maximum from direct optimization. Open markers are the five measured runs with settings drawn independently for each cell, one color and symbol per processor, namely \texttt{ibm\_fez} (green circles,
		$m=8$ and $m=16$), \texttt{ibm\_berlin} (purple square, $m=8$) and \texttt{ibm\_marrakesh} (orange
		triangles, $m=6$ and $m=8$), with $40$ configurations of $8192$ shots except the $m=16$ run, which used
		$400$ configurations of $800$ shots. Points at $m=8$ are offset horizontally by $\pm0.16$ for legibility.
		Statistical errors, the standard error over configurations, are $0.001$ to $0.005$ and smaller than the
		markers. The measurements cover $m=6$, $8$ and $16$, and the curves beyond that are the theoretical bounds.}
	\label{figmono}
\end{figure}

Eight cells on $48$ disjoint qubits of \texttt{ibm\_fez}, with each cell's settings drawn independently over
$40$ configurations, all certify individually and win jointly with probability $0.503\pm0.002$ against
$\beta^8=0.282$ for any Gaussian strategy and $0.75^8=0.100$ for any $\NC$ circuit (Fig.~\ref{figmono}). Every
configuration certifies on its own, and the product of the single cell win probabilities, $0.502$, matches
the joint value, so the cells are independent. The same test gives $0.457\pm0.002$ for eight cells on
\texttt{ibm\_berlin} and $0.645\pm0.003$ for six on \texttt{ibm\_marrakesh}, against $\beta^6=0.387$. Sixteen
cells on $96$ qubits of \texttt{ibm\_fez}, with $400$ independent configurations of $800$ shots, win jointly with
probability $0.174\pm0.001$ against $\beta^{16}=0.079$, a ratio of $2.2$, the worst configuration at $0.134$.
Thirteen of the sixteen cells certify on their own and three do not, and the joint result does not depend on
that, which is what a monogamy law means. The empirical Bernstein inequality certifies this run against any
Gaussian strategy with no assumption on the noise at confidence $1-10^{-4}$, and at $1-10^{-6}$ by a smaller
margin (Methods). On processors where the layouts extended into poorer regions the dual rail acceptance of
each cell tracked its Mermin value, with correlations of $+0.81$ to $+0.97$, so the flag recorded in every shot
at no cost identifies the cells whose value is degraded.

\begin{figure}[t]
	\centering
	\includegraphics[width=\linewidth]{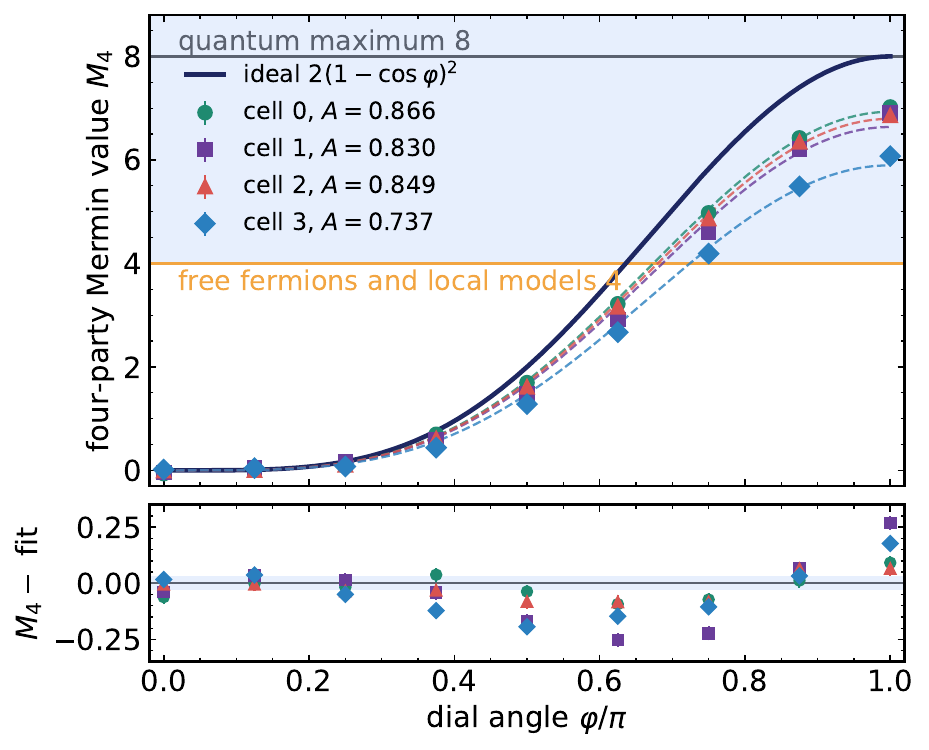}
	\caption{\textbf{Four parties.} The Mermin value $M_4$ of four star cells on $32$ qubits of \texttt{ibm\_fez},
		$8192$ shots per setting, raw data, against the dial angle. The solid curve is the ideal
		$2(1-\cos\varphi)^2$ and the dashed curves are single visibility fits per cell. The free fermion bound of
		Theorem~3 at $n=4$ is $4$, which is also the local bound, so the shaded region is a Bell violation that
		certifies magic. The lower panel shows the residuals to the fits against the shot noise band.}
	\label{figstar}
\end{figure}

Theorem~3 at four parties was tested with a star cell, one centre domino coupled to three leaf dominoes
(Methods). A chain of four would give a linear cluster state whose Mermin value is exactly the free fermion
bound $4$ and would certify nothing. The star's ideal signal is $M_4(\varphi)=2(1-\cos\varphi)^2$, reaching the
quantum maximum $8$ and crossing the bound $4$ at the same angle $114.47^\circ$ as the cell of three parties, since
$2(1-\cos\varphi)^2=4$ and $2(1-\cos\varphi)=2\sqrt2$ are the same equation. Four cells on $32$ qubits of
\texttt{ibm\_fez} give $M_4=7.02$, $6.91$, $6.86$ and $6.07$ at $\varphi=\pi$, each $\pm0.03$, against the bound
$4$, which at four parties is also the local bound, with every cell certifying from $\varphi=0.75\pi$ on
(Fig.~\ref{figstar}). The eight signed correlators of each cell agree to within $0.06$, so the loss is a
single visibility, $0.74$ to $0.87$, and the measured crossings sit where each cell's visibility predicts them
to within four degrees.

\subsection*{Fermionic matter and a phase transition}

Fermionic Gaussian states are the states of every quadratic fermionic Hamiltonian, which includes the BCS
ground state of a conventional superconductor, the Kitaev chain, every Slater determinant and every
Hartree-Fock state, together with their thermal states. Theorems~1 to~3 hold for all of them. Under local
fermionic readout of the kind they cover, mean field fermionic matter never exceeds the biseparable Mermin
bound and is never genuinely multipartite nonlocal, however strongly it is paired. The obstruction is fermion
parity, which every physical fermionic Hamiltonian conserves. Read in reverse, a Mermin value above the bound
in a fermionic system certifies correlations that no quadratic Hamiltonian can produce, a witness of
interaction beyond mean field, and the readout it requires, mode resolved occupation after local Gaussian
rotations, is native to quantum dot arrays, Majorana devices and fermionic quantum gas microscopes.

\begin{figure}[t]
	\centering
	\includegraphics[width=\linewidth]{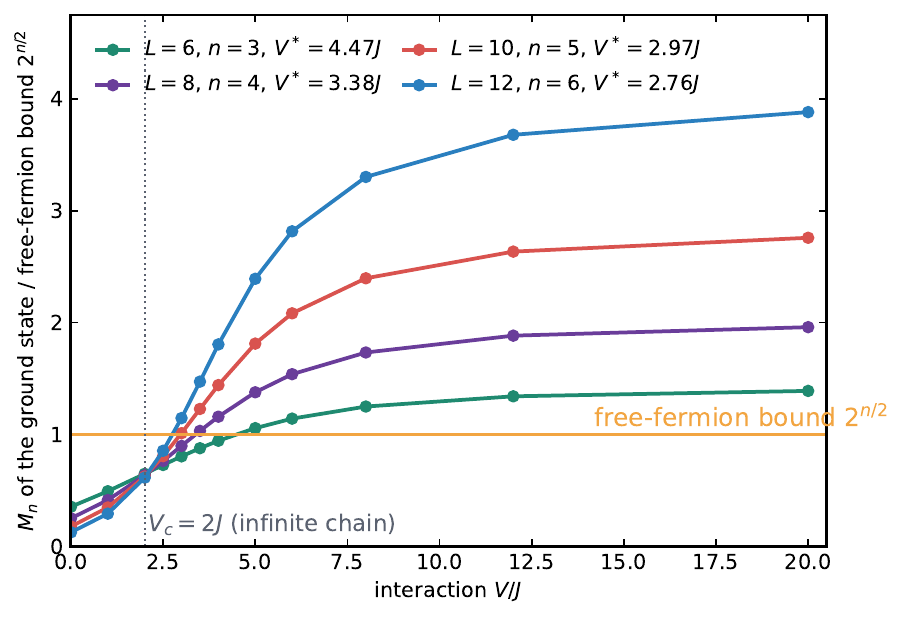}
	\caption{\textbf{The witness at the charge density wave transition.} Ground state Mermin value of the ring
		of Eq.~\eqref{eqtV}, maximized over logical measurement settings of the $n$ site pairs and divided by the
		free fermion bound $2^{n/2}$ of Theorem~3, against the interaction $V$, for rings of $6$ to $12$ sites. At
		$V=0$ the ground state is Gaussian and sits below the bound. It crosses the bound at $V^*(L)$, which moves
		toward the infinite chain transition $V_c=2J$ as the ring grows, and saturates at the cat value
		$2^{n-1}$, so the ratio to the bound grows as $2^{n/2-1}$ with system size.}
	\label{figcdw}
\end{figure}

The witness can be put to a physical model.Consider spinless fermions on a ring of $L=2n$ sites at half filling,
\begin{equation}
H=-J\sum_{i}\bigl(c_i^\dagger c_{i+1}+\mathrm{h.c.}\bigr)+V\sum_i n_in_{i+1},
\label{eqtV}
\end{equation}
which is the XXZ chain in disguise and passes from a Luttinger liquid to a charge density wave at $V_c=2J$
in the thermodynamic limit. Group the sites into $n$ adjacent pairs and treat each pair as a dual rail party,
so that every measurement is a logical observable, the settings share a direction automatically, and
Theorems~1 to~3 apply with no restriction. At $V=0$ the ground state is Gaussian and Theorem~3 caps its
Mermin value of $n$ parties at $2^{n/2}$. Deep in the CDW phase the finite ring ground state is the cat
$(\ket{1010\cdots}+\ket{0101\cdots})/\sqrt2$, which in the pair language is $\ket{00\cdots}_L+\ket{11\cdots}_L$,
a GHZ state of the $n$ parties with Mermin value $2^{n-1}$. The ground state must therefore cross the
free fermion bound at some interaction $V^*(L)$, and exact diagonalization locates it (Fig.~\ref{figcdw}).

Three things come out. The onset $V^*$ is $4.47J$, $3.38J$, $2.97J$ and $2.76J$ for $6$, $8$, $10$ and $12$
sites, moving toward $V_c=2J$ as the ring grows. Nonlocality beyond free fermions is a finite size marker
of the transition, placed inside the ordered phase where the two density patterns have formed a coherent
superposition. Second, the ratio of the cat's Mermin value to the free fermion bound is $2^{n/2-1}$, so the
visibility at which a noisy measurement still certifies falls from $0.72$ at three parties to $0.26$ at
six, and continues to halve with every four sites. A larger ring tolerates more noise, the opposite of an
ordinary multipartite Bell test, because the obstruction being tested, the free fermion bound, grows only
as the square root of the signal. Third, boundaries matter. On an open chain the particles can avoid each
other in more than two patterns and single hops connect them, so the strong coupling ground state is not a
cat and its Mermin value never reaches the local bound. The witness sees the ring's cat and not the chain's
spread, which is the correct physics.

\begin{figure}[t]
	\centering
	\includegraphics[width=\linewidth]{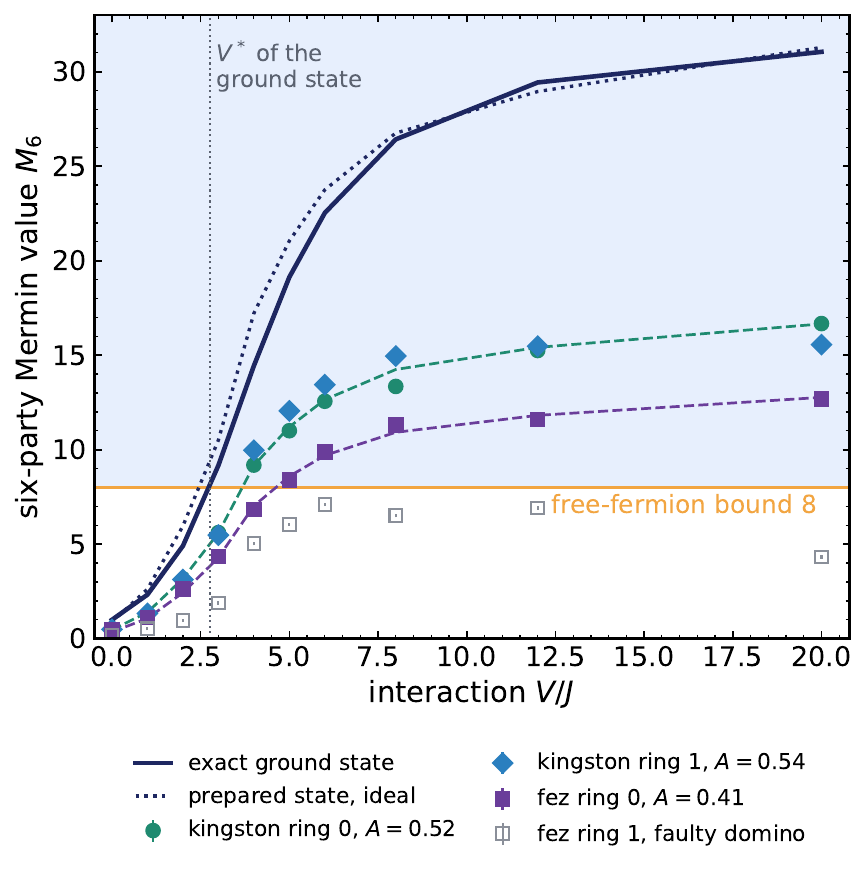}
	\caption{\textbf{The interacting fermion ring on hardware.} Mermin value of six parties of the twelve site ring of
		Eq.~\eqref{eqtV} prepared on twelve qubit loops of two processors, against the interaction $V$, raw data,
		$32$ settings of $4096$ shots per point. The solid curve is the exact ground state and the dotted curve the
		ideal prepared state. Filled markers are the measured loops, two on \texttt{ibm\_kingston} and one on
		\texttt{ibm\_fez}, each following the prepared curve times a single visibility (dashed, the mean for each
		processor) and crossing the free fermion bound $8$ where that visibility predicts. Open squares are a
		second \texttt{ibm\_fez} loop that contained one faulty domino. The dotted vertical line marks the crossing of
		the exact ground state at $2.76J$. Error bars are the shot noise, $0.09$, smaller than the markers.}
	\label{figringhw}
\end{figure}

The cells of four parties measured above are this model's strong coupling limit. The star circuit at
$\varphi=\pi$ prepares, up to local Cliffords, the eight site cat, whose ground state value at $V=20J$ is
$7.84$, and the hardware gave $7.02$. We measured the twelve site ring directly, on two processors. A twelve qubit hexagonal loop of a heavy hex chip
carries the ring with no swaps, and its closing bond needs no extra sign at fixed particle number. Starting from
the domino superpositions used in the cells, three layers of the model's own gates, controlled phases for the
interaction and $XX+YY$ rotations for the hopping, prepare the ground state with fidelity $0.84$ at $V=0$, $0.89$ to
$0.96$ near the onset and $0.99$ at strong coupling, in $114$ two qubit pulses (Methods). One fixed set of logical
Pauli settings, optimal for the strong coupling cat, is also optimal for the ground state at every $V$, so nothing
is tuned point by point. Figure~\ref{figringhw} shows the Mermin value of six parties at ten interaction strengths,
each from $32$ settings of $4096$ shots, for two loops of \texttt{ibm\_kingston} and two of \texttt{ibm\_fez}. On
every loop without a fault the measured curve is the predicted curve of the prepared state rescaled by a single
visibility, to within about ten percent at every $V$ from $J$ up, $A=0.52$ and $0.54$ on \texttt{ibm\_kingston} and $0.41$ on \texttt{ibm\_fez}. Each stays below the
free fermion bound at weak interaction and crosses it, at $V=3.67J$ and $3.56J$ on \texttt{ibm\_kingston} and $4.74J$
on \texttt{ibm\_fez}, where the visibilities predict $3.71J$, $3.64J$ and $4.62J$. The crossing is set by the
ground state, at $2.76J$, shifted by the visibility to where the prepared value reaches $8/A$, and it approaches the
ground state's as the hardware improves. Above it the values exceed the bound by up to $8.7$ against a shot noise of
$0.09$. Hoeffding's inequality on the full sum of shots, which assumes nothing about the noise, certifies every
point from $V=4J$ on both \texttt{ibm\_kingston} loops at confidence $1-10^{-6}$, and from $V=6J$ on
\texttt{ibm\_fez}. One \texttt{ibm\_fez} loop contained a domino whose dual rail flag passed $54\%$ of shots against
$80\%$ to $93\%$ elsewhere, and it certified at no $V$. The flag identifies the faulty domino without reference to
the Mermin value, as it identified the weak cells of the multicell runs.

\subsection*{Outlook}

For direction sharing settings the free fermion value of the Mermin family is now known completely, attained
by explicit states, and measured at three parties, at four, on 96 qubits, and across an interaction driven
transition on two processors. What remains open is stated in Methods, the constant for arbitrary settings, which
every numerical search puts at $2\sqrt2$ while the certified bound is $3.26$, the angle resolved law beyond
three parties, the thermodynamic limit of the transition witness, and the connection between a measured
violation and the classical cost of reproducing the data. The broader lesson is that the boundary of
classical simulability is not one thing. For Clifford circuits it leaves maximal nonlocality on the
simulable side. For fermions it is drawn exactly at the point where a system of many parties begins to carry
more Bell nonlocality than one pair, and a quantum processor can watch that boundary being crossed.

\section*{Methods}

\subsection*{Setting}

Let $c_1,\dots,c_N$ be Majorana operators, $\{c_a,c_b\}=2\delta_{ab}$. A fermionic Gaussian state, pure or
mixed, is fixed by its covariance matrix $\Gam_{ab}=\tfrac{i}{2}\avg{[c_a,c_b]}$, a real antisymmetric
contraction, with all higher correlators given by Wick's theorem, $\avg{g_1\cdots g_{2m}}=\Pf(G)$ for linear
forms $g_i=w_i\!\cdot\!c$ and $G_{ij}=w_i\!\cdot\!w_j-i\,w_i^{\top}\Gam w_j$ for $i<j$. All statements extend by
linearity to convex mixtures of Gaussian states. Bilinear readout respects fermionic superselection but is a
genuine restriction, because products of commuting bilinears, the quartic observables used for the
pseudotelepathy of ref.~\cite{CampbellHobanEisert2014}, are also parity preserving.

\subsection*{Proof of Theorem 1}

\emph{Reduction.} By multilinearity of the Pfaffian, the value depends on $\Gam$ only through its
compression onto the $4n$ frame directions, which is the covariance of a mixed Gaussian state on $2n$ modes.
The bound is therefore independent of the number of modes per party, and it suffices to treat $k=2$.

\emph{An inequality for vacuum and full occupation.} For any mixed Gaussian state on $M$ modes and any occupation pattern $n$,
\begin{equation}
P(n)P(\bar n)\le
\begin{cases} 2^{-M}, & M\ \text{even},\\ 2^{-M-1}, & M\ \text{odd},\end{cases}
\label{eqvf}
\end{equation}
both tight. This follows from $P(n)P(\bar n)=4^{-M}\sqrt{\det(1-S^{2})}$ with $S=\Gam\Gam_n$. Since
$\Gam_n^{-1}=-\Gam_n$, the characteristic polynomial of $S$ is the square of a real polynomial, so the
spectrum of $S$ is doubly degenerate and, for odd $M$, contains a real eigenvalue. Parity of the mode number
is what forbids the paradox for three parties.

\emph{Helper modes.} One has the operator identity $Q_j^{\dagger}Q_j=2(1+s_j\tilde Z_j)$ with $\tilde Z_j$ a
bilinear anticommuting with both settings and $s_j=|\sin\varphi_j|$. In the joint eigenbasis of the three
$\tilde Z_j$, $\avg{Q_AQ_BQ_C}=\sum_{k,\epsilon}\bigl(\prod_j q^{(j)}_{\epsilon_j}\bigr)
\langle\bar n_{k\epsilon}|\rho|n_{k\epsilon}\rangle$, and positivity gives
$|\langle\bar n|\rho|n\rangle|\le\sqrt{P(n)P(\bar n)}$. Applying Eq.~\eqref{eqvf} with $M=3$, together with
$\sum_{\text{pairs}}\sqrt{P P'}\le\tfrac12$, reduces the problem to maximizing a linear form over a polytope,
whose value is $\tfrac{\sqrt2}{2}[V_{(1)}+V_{(2)}]$ with
\begin{equation}
V(\epsilon)=\prod_j\sqrt{1+\epsilon_js_j}+\prod_j\sqrt{1-\epsilon_js_j}
\label{eqV}
\end{equation}
and $V_{(1)}\ge V_{(2)}$ its two largest values on $\{\pm1\}^3$ modulo a global sign. The elementary
inequality $V(\epsilon)+V(\epsilon')\le4$ for noncomplementary $\epsilon,\epsilon'$ closes it. For
$s_j\equiv1$ only one weight survives and the bound falls to $2$. $\square$

\emph{Arbitrary planes.} Without the hypothesis each party carries two commuting helper bilinears, and the
same construction applied to every submarginal yields a linear program in $32$ pair weights. The relaxation
is rigorous at every fixed angle vector. Its maximum over the six angles is located numerically at a
corner, $2\sqrt2\,(\tfrac34+\tfrac{1}{2\sqrt2})=3.12$, and interval branch and bound certifies
$|\avg{Q_AQ_BQ_C}|\le3.26$ for all bilinear settings. Genuine tripartite pseudotelepathy is therefore
impossible under \emph{any} bilinear settings. The sharp constant, and with it the Svetlichny corollary, is
proven under the hypothesis of Theorem~1, since $3.26$ alone would allow $S_3$ up to $4.61$. Numerically the
hypothesis is not needed. Local maximization over pure and mixed states with $k=2,3,4$ modes per party and
arbitrary plane angles returns $2.828427$ in every run, and for $k=2$ letting each party read all occupation
numbers with any of the $16$ Boolean postprocessings per setting, all $16^{6}$ assignments enumerated at
each step of a local search, gives no improvement (Supplementary Information).

\subsection*{Proof of Theorem 2}

\emph{Proof.} The helper mode argument of Theorem~1 stops, before its final inequality, at
$|\avg{Q_AQ_BQ_C}|\le\tfrac{\sqrt2}{2}[V_{(1)}+V_{(2)}]$. Because $\epsilon$ and $-\epsilon$ are identified,
two distinct classes have representatives differing in exactly one position $p$, and then
\begin{multline*}
V(\epsilon)+V(\epsilon')=\Bigl[\sqrt{1+s_p}+\sqrt{1-s_p}\Bigr]\\
\times\Bigl[\sqrt{(1+\epsilon_qs_q)(1+\epsilon_rs_r)}+\sqrt{(1-\epsilon_qs_q)(1-\epsilon_rs_r)}\Bigr].
\end{multline*}
With $\sqrt{1\pm\sin\varphi}=\cos\tfrac{\varphi}{2}\pm\sin\tfrac{\varphi}{2}$ on $[0,\pi/2]$, the first bracket is
$2\cos\tfrac{\varphi_p}{2}$ and the second is $2\cos\tfrac{\varphi_q-\varphi_r}{2}$ when $\epsilon_q=\epsilon_r$ and
$2\cos\tfrac{\varphi_q+\varphi_r}{2}$ otherwise, the former being the larger. This gives the upper bound
$B(\varphi)$. For attainment, write $Q_p=ic_1^p\ell_p$ with $\ell_p=(1+i\cos\varphi_p)c_2+i\sin\varphi_p\,c_3$.
Pair parties $q$ and $r$ perfectly, $c_1^q$ with $c_1^r$, $c_4^q$ with $c_4^r$, and the $(c_2,c_3)$ plane of
$q$ with that of $r$ through the orthogonal map $\mathrm{Rot}(\tfrac{\varphi_q+\varphi_r}{2})\,
\mathrm{diag}(1,-1)$. Wick's theorem leaves one contraction and gives
$|\avg{Q_qQ_r}|=2\cos\tfrac{\varphi_q-\varphi_r}{2}$. Give party $p$ the pure state pairing $c_1$ with the
bisector $\cos\tfrac{\varphi_p}{2}c_2+\sin\tfrac{\varphi_p}{2}c_3$ of its two settings and $c_4$ with the
orthogonal direction, which gives $|\avg{Q_p}|=\sqrt2\cos\tfrac{\varphi_p}{2}$, the maximum for one party. The
Pfaffian factorizes on the product state and the two values multiply. Since each factor is at most one,
$B=2\sqrt2$ requires $\varphi_p=0$ and $\varphi_q=\varphi_r$. $\square$

The theorem connects the two cases of Theorem~1, $\varphi=0$ and $\varphi=(\tfrac\pi2,\tfrac\pi2,\tfrac\pi2)$,
and says what lies between them. The free fermion value is largest not at one point but on a family, one
party passive and the other two at any common angle, and it degrades smoothly as the angles separate. The
party that is left alone is the one whose planes are closest, and the other two are paired. A search over
a $61^3$ grid finds the minimum $B=2$, the local bound, at exactly four settings, all three anticommuting
or one anticommuting and two identical. Obtuse angles reduce to the range of the theorem by relabeling one
outcome, which conjugates that party's $Q_p$, and the upper bound then holds with $\varphi_p$ replaced by
$\arcsin|\sin\varphi_p|$. Direct numerical maximization attains it in every such case tested, including mixed
acute and obtuse settings. Direct maximization at $32$ canonical settings agrees with Eq.~\eqref{eqBphi} to
five decimals, and the constructed states reproduce it to machine precision at $505$ settings (Supplementary Information). The
Mermin polynomial itself, which is one phase of the correlator, is in general strictly below $B(\varphi)$ at
fixed settings and reaches $2\sqrt2$ only at the Tsirelson point $\varphi=(\tfrac\pi2,\tfrac\pi2,0)$ and its
permutations, since the construction fixes the modulus of the correlator but not its phase.

\subsection*{Proof of Theorem 3}

\emph{Proof.} Write $Q_j=i\,a_jb_j$ with the linear forms $a_j=u_j\!\cdot c$ and $b_j=\ell_j\!\cdot c$,
$\ell_j=v_j^0+iv_j^1$. The parties' operators commute, so $\prod_jQ_j=i^n\,a_1b_1\cdots a_nb_n$ is an ordered
product of $2n$ linear forms, and Wick's theorem gives its expectation as $i^n\,\Pf(G)$. Every symmetric part
of $G$ vanishes, since $u_j\!\cdot\ell_j=0$ and forms of different parties have disjoint support, so
$G=-iW^{\!\top}\Gam W$ with $W=[u_1,\ell_1,\dots,u_n,\ell_n]$, and
\[
\Bigl\langle\prod_jQ_j\Bigr\rangle=\Pf\bigl(W^{\!\top}\Gam W\bigr).
\]
For an antisymmetric matrix $\Pf^2=\det$. The Cauchy-Binet expansion with Cauchy-Schwarz over row subsets
gives $|\det(X^\dagger Y)|^2\le\det(X^\dagger X)\det(Y^\dagger Y)$, and with $X=\overline W$, $Y=\Gam W$ and
$\Gam^{\!\top}\Gam\le1$,
\[
\Bigl|\Bigl\langle\prod_jQ_j\Bigr\rangle\Bigr|^2=\bigl|\det(W^{\!\top}\Gam W)\bigr|\le\det(W^\dagger W)=2^n,
\]
because $W^\dagger W$ is block diagonal over parties with blocks $\mathrm{diag}(\|u_j\|^2,\|\ell_j\|^2)=
\mathrm{diag}(1,2)$. On a product state the Pfaffian factorizes over the blocks, each pair at equal angles
contributes $2$ and a lone party at angle zero contributes $\sqrt2$. $\square$

For the Mermin polynomial and every $n\ge4$ the value $2^{n/2}$ lies strictly below the biseparable quantum
bound $2^{n-3/2}$, so the coincidence with the biseparable bound is special to three parties. For arbitrary
planes $Q_j$ no longer factors into two linear forms and the law remains a conjecture, verified numerically at
$n=3,4,5$.

\subsection*{Proof of Theorem 4 and the circuit statement}

\emph{Proof.} With the four inputs drawn uniformly a cell wins with probability $\tfrac12+\avg{M_3}/8$, so
Theorem~1 gives $\beta=\tfrac12+2\sqrt2/8$ for one cell. Fix the settings. Each observable is the occupation
number of one rotated mode, and the $3m$ measured modes are orthogonal, so all observables commute.
Conditioning a Gaussian state on occupation outcomes leaves a Gaussian state on the remaining modes, and
conditioning on a win event, a union of outcomes, leaves a convex mixture of them. The settings of cell $i$
are independent of everything conditioned on, so $\Pr[W_i\mid W_1\cdots W_{i-1}]=\tfrac12+\avg{M_3}_i/8\le\beta$
by Theorem~1 on the conditional state, and the bound multiplies. $\square$

Gaussianity is preserved by exactly the conditioning the induction requires, so a Gaussian strategy cannot
bank correlations in one cell to pay for another. The same induction gives nothing for general quantum
states. Direct maximization for $m=2$, one global mixed Gaussian state on six parties with unrestricted
covariance between the cells, $336$ parameters and exact Pfaffian correlators, converges to $\beta^{2}=0.728553$
from random initializations and never exceeds it (Supplementary Information).

Theorem~4 converts into a circuit statement. For a constant depth circuit of nearest neighbour matchgates
with a fixed Jordan-Wigner ordering, acting on an input in the computational basis, with the three parties' input bits
loaded at qubits with pairwise disjoint forward light cones, the output covariance factorizes as
$\Gam(s)=M_A(s_A)M_B(s_B)M_C(s_C)\,\Gam_0\,(\cdot)^{\top}$ with each $M_p$ orthogonal and supported in party
$p$'s cone. The bit flip is the reflection $\mathrm{diag}(1,-1)$ on one Majorana block, and since $X_p$
carries a Jordan-Wigner string the statement holds at the level of the covariance (Supplementary Information). Each party thus sees
one fixed Gaussian state, an orthogonal transformation in its cone that depends on the setting, and occupation readout.
If each party outputs one designated qubit, its observable is a bilinear in its cone and Theorem~4 applies
with $\beta'$, or with $\beta$ under the hypothesis of Theorem~1. Multiqubit readout with postprocessing is
covered only by the numerics above.

Consequently there is a relational problem, defined by $m$ Mermin cells at pairwise distance exceeding twice
the circuit depth, that a constant depth circuit of matchgates plus controlled phase gates solves with
probability $1$, while every constant depth matchgate circuit succeeds with probability at most
$\beta'^{\,m}$, or $\beta^m$ under the hypothesis of Theorem~1, and every $\NC$ circuit at most
$0.75^{m}$~\cite{Mermin1990,BGK2018}. Both families are geometrically local, constant depth and classically
simulable, so the separation is in correlation power, and they differ only in the phase of one gate. The
separating resource is fermionic magic. This is the free fermion counterpart of the Clifford plus magic
theorem on shallow circuits of Ref.~\cite{Zhang2024}, with an exponentially rather than inverse polynomially small
success probability, and it is not implied by the restriction of matchgates to linear threshold
functions~\cite{VandenNest2011}, which concerns single output functions at unrestricted depth.

\subsection*{The certificate and its statistics}

\emph{A hardware criterion.} At the operating point $\varphi=\pi$ the ideal signal is $M=4$, and incoherent
noise rescales it by a visibility $A$, so the measured value is $4A$. Certification requires
\begin{equation}
A>\frac{1}{\sqrt2}\approx0.707 .
\label{eqvis}
\end{equation}
The same threshold governs monogamy. A single cell wins with probability $\tfrac12+A/2$, which exceeds the
Gaussian ceiling $\beta$ precisely when $A>1/\sqrt2$, and at four parties the threshold drops to $A>1/2$, so a set of cells that each exceed the threshold passes
the multicell test for every $m$. Certification at $k$ standard deviations needs about
$\bigl(2k/(4A-2\sqrt2)\bigr)^2$ shots per setting, which is $401$ shots for $5\sigma$ at the visibility
$A=0.832$ measured on \texttt{ibm\_fez}, and $238$ at $A=0.869$ on \texttt{ibm\_berlin}.

\emph{Without a Gaussian approximation.} The estimator $\hat M$ is a signed sum of four correlators, each the
mean of $N$ outcomes in $\{\pm1\}$, so Hoeffding's inequality gives, with probability at least $1-\delta$,
\begin{equation}
|\hat M-M|\le4\sqrt{\frac{2\ln(8/\delta)}{N}} .
\end{equation}
With $N=8192$ this is $0.249$ at $\delta=10^{-6}$ and $0.298$ at $\delta=10^{-9}$, so any measured $\hat M>3.13$
certifies non-Gaussianity at confidence $1-10^{-9}$ with no assumption on the noise. For the monogamy
statistic the same inequality applies at the level of configurations, where the success
fractions lie in $[0,1]$, with margin $\sqrt{\ln(2/\delta)/(2n)}$ for $n$ configurations. A test designed for a
certificate that assumes nothing about the noise should therefore spend its budget on many configurations with fewer shots each,
which is how the run with sixteen cells was taken, $400$ configurations of $800$ shots. Its joint value sits $0.095$
above $\beta^{16}$, below the Hoeffding margin of $0.135$ at $\delta=10^{-6}$, because the joint probability is
small at sixteen cells. The empirical Bernstein inequality, which uses the measured spread of $0.016$ across
configurations rather than the worst case, certifies it at confidence $1-10^{-4}$ with $0.113$ against $0.079$,
and at $1-10^{-6}$ with $0.085$.

\emph{A bound on a magic monotone.} Theorem~1 holds on the convex hull of Gaussian states, so resource
monotones defined by convex decompositions inherit a bound from the data. For the Gaussian robustness $R$,
the minimal $\sum_i|q_i|$ over decompositions $\rho=\sum_iq_i\sigma_i$ into Gaussian states, the fermionic
analogue of the robustness of magic~\cite{HowardCampbell2017,Brandao2005,TakagiRegula2019}, every expectation
value obeys $|\avg{O}_\rho|\le R\max_{\rm Gaussian}|\avg{O}|$. The measured values therefore certify, with no
knowledge of the state, $R\ge M_{\rm raw}/2\sqrt2$, which is $1.17$ on \texttt{ibm\_fez} and $1.25$ on
\texttt{ibm\_berlin} for the six qubit states, and from the joint success probability,
$\Pr[\text{all }m\text{ win}]\le R\,\beta^m$, $R\ge1.79$ for the state of $48$ qubits of \texttt{ibm\_fez}, with
$1.62$ and $1.67$ on the other two devices. These bounds concern a resource measure and not simulation cost,
since the best matchgate plus magic simulators scale with the extent~\cite{ReardonSmith2024,DiasKoenig2024}.

\subsection*{The dual rail cell}

Encode each logical qubit in two
adjacent physical qubits, a dual rail with code space $\{\ket{01},\ket{10}\}$, so that $Z_L=Z_{r_0}$,
$X_L=X_{r_0}X_{r_1}$ and $Y_L=Y_{r_0}X_{r_1}$ are Majorana bilinears, and every single qubit logical Clifford
and every logical Pauli readout, a matchgate rotation followed by $Z_{r_0}$, is Gaussian. The single
non-Gaussian element is the controlled phase between dominoes $\mathrm{CZ}_L(\varphi)=\mathrm{CP}(\varphi)=
\mathrm{fSim}(0,\varphi)$, a matchgate if and only if $\varphi\equiv0$, so $\varphi$ is a continuous
fermionic magic dial. In fermionic terms $\mathrm{CP}(\varphi)=e^{i\varphi\,n_rn_{r'}}$ on one rail of each
domino, an interaction between the densities of two modes applied for a time $\varphi$ in units of its
strength, while every other gate is quadratic. The dial is therefore the interaction, and the thresholds
below are thresholds in interaction strength. Since $Z_L=Z_{r_0}=-Z_{r_1}$ on the code space the controlled phase may act on either
rail, so a cell is a bare line of six qubits (Fig.~1) with three two qubit layers and no swaps,
embeddable on any heavy hex processor. With fractional gates the dial is one pulse,
$\mathrm{CP}(\varphi)=e^{i\varphi/4}R_z(\tfrac{\varphi}{2})R_z(\tfrac{\varphi}{2})R_{zz}(-\tfrac{\varphi}{2})$.

With settings $(Y,Z)$, $(X,Y)$, $(Y,Z)$ on the three parties and the outcome of party $C$'s first setting
relabeled to $-Y_L$, one of four optimal assignments found by exhaustive search over all $216$ pairs of Paulis per party, the measured polynomial is
\begin{multline}
  M=-\avg{Y_AX_BY_C}-\avg{Y_AY_BZ_C}\\
  -\avg{Z_AX_BZ_C}+\avg{Z_AY_BY_C},
\end{multline}
and the ideal signal is exactly
\begin{equation}
  M(\varphi)=2\,(1-\cos\varphi),
\end{equation}
so the thresholds are analytic. The local bound is crossed at $\varphi=\pi/2$ and the Gaussian bound at
\begin{equation}
  \varphi^{\ast}=2\arcsin 2^{-1/4}=1.997875\ \text{rad}=114.47^{\circ},
\end{equation}
beyond which the data ceases to be reproducible by any free fermion circuit with local bilinear readout
(Fig.~\ref{figdial}). Magic is present for every $\varphi\neq0$, and $\varphi^{\ast}$ is where it becomes
certifiable without trusting the measurement angles. Trusting that the Pauli settings anticommute exactly,
Theorem~2 gives the bound $2$ and certification begins at $\varphi=\pi/2$. Every logical measurement of a dual rail qubit is a bilinear in $c_1,c_2,c_3$, so the
compiled settings satisfy the hypothesis of Theorem~1 exactly.

We ran the protocol for three parties on three IBM processors of two architectures, \texttt{ibm\_fez} and
\texttt{ibm\_marrakesh} (Heron~r2) and \texttt{ibm\_berlin} (Nighthawk~r1), with $8192$ shots per setting,
dynamical decoupling and measurement twirling, and no readout mitigation or extrapolation of any kind. All
twelve runs performed for this work are listed in the SM, including those not quoted here.

\emph{The dial.} One cell on \texttt{ibm\_fez}, swept over nine values of $\varphi$, gives a raw Mermin value
rising monotonically to $M_{\rm raw}=3.305\pm0.022$ at $\varphi=\pi$, $0.48$ above $2\sqrt2$, the uncertainty
being the worst case shot noise $2/\sqrt N$ (Fig.~\ref{figdial}). Three independent jobs at $\varphi=\pi$ gave $3.340$,
$3.243$ and $3.305$, all certifying, the weakest by $0.41$. A fit with one parameter $M=A\,2(1-\cos\varphi)$
returns a visibility $A=0.832\pm0.004$ independent of $\varphi$, with residuals at the shot noise level, so the
measured curve is the predicted cosine rescaled, with no distortion of its shape. The same scan on
\texttt{ibm\_berlin} gives $M_{\rm raw}=3.538\pm0.022$ at $\varphi=\pi$, $0.71$ above $2\sqrt2$, with
$A=0.869\pm0.005$, and the same line measured again in the run with eight cells below gave $3.517$. A repeat compiled
without fractional gates gives $3.087$, so certification does not depend on the gate decomposition. Both
single cell values also certify under the bound that assumes nothing about the noise, at confidence $1-10^{-9}$.

Against the certified bound $3.26$ for arbitrary settings, which assumes nothing about the measurement planes, the \texttt{ibm\_berlin} single cell certifies by $0.28$, as do the monogamy results with six cells and with eight cells on \texttt{ibm\_fez} and the best cells on all three devices (Supplementary Information). Postselection
is not free for this bound. A Gaussian strategy discarding the same fraction of rounds reaches
$T=\max\{C/\alpha_{\min},\,C+3(1-\alpha_{\min})\}$ with $C=2\sqrt2+\sum_s(1-\alpha_s)$ (Supplementary Information), so flagged data must
exceed $T$, not $2\sqrt2$. On \texttt{ibm\_fez} the flagged value never clears $T$ with confidence in any of six single cell
runs at $\varphi=\pi$. With fractional gates it lies within $0.06$ of $T$, above it in two runs and below in one,
never by more than two standard errors, and without them it falls $0.33$ to $0.40$ below. On \texttt{ibm\_berlin},
whose dual rail acceptance is higher, the threshold falls to $3.539$ and the flagged value $3.917$ clears it by
$0.38$, more than the margin that assumes nothing about the noise at confidence $1-10^{-6}$. The naive comparison with $2\sqrt2$
would have passed by about $1.0$ on every run. We quote only the raw value, which needs no correction and
certifies on every device, in which each party reports its rail-$r_0$ bit whether flagged or not. That bit is a bilinear, so the
raw value is exactly what Theorem~1 bounds. The statistical uncertainties are ten to a hundred times smaller
than the margins, so what matters is systematics. Gate errors and decoherence only reduce the raw correlators.
Independent readout errors are classical postprocessing of each party's own outcome, a mixture of reporting
$\pm O$ or a fixed value, which is itself a Gaussian strategy and cannot produce a violation. Drift between runs
is $0.05$, a tenth of the single cell margin. What lies outside the model is readout crosstalk between parties,
at the percent level on these devices and far below any margin quoted here. Readout mitigation and zero noise
extrapolation rescale
correlators, are unavailable to a Gaussian strategy, and were not applied.

\subsection*{The star cell}

Theorem~3 at $n=4$ was tested with a star cell, one center domino coupled by
controlled phases to three leaf dominoes, eight qubits in all (Fig.~4). The star is forced. Extending
the line to four dominoes gives a linear cluster state whose Mermin value of four parties is exactly $4$ for every
logical Pauli setting, the free fermion bound itself, so it would certify nothing even without noise. On a
heavy hex processor the center must couple two leaves through one rail and the third through the other, and
the settings, $(X,Y)$, $(Y,Z)$, $(Y,Z)$ and $(Z,Y)$ with the last party's outcomes negated, were found by
exhaustive search for that circuit. Its ideal signal is
\begin{equation}
M_4(\varphi)=2(1-\cos\varphi)^2,
\end{equation}
reaching the quantum maximum $8$ at $\varphi=\pi$ and crossing the bound $4$ at $\varphi^*=114.47^\circ$, the
same angle at which the signal of three parties crosses $2\sqrt2$, since $2(1-\cos\varphi)^2=4$ and
$2(1-\cos\varphi)=2\sqrt2$ are the same equation. The certification threshold on the dial is therefore
common to both cells. At $\varphi=\pi$ a visibility above $1/2$ suffices, against $1/\sqrt2$ for three parties.
Four cells on $32$ disjoint qubits of \texttt{ibm\_fez}, each swept over the same nine angles with $8192$ shots
per setting, give $M_4=7.02$, $6.91$, $6.86$ and $6.07$ at $\varphi=\pi$, each $\pm0.03$, against the bound $4$,
with every cell certifying from $\varphi=0.75\pi$ on. The eight signed correlators of each cell agree to within
$0.06$, so the loss is a single visibility, $A=0.87$, $0.83$, $0.85$ and $0.74$, within the range of the
cells of three parties although the circuit is a third longer. The weakest cell traces to one leaf domino whose
dual rail flag passes $91\%$ of shots against $95\%$ to $97\%$ for the others. The measured crossings of the
bound, $122^\circ$ to $132^\circ$, sit where each cell's visibility predicts them to within four degrees. Two
cells show residuals to the single visibility curve of up to $0.27$ with a monotone shape, steeper than the
ideal in the middle of the dial, which lowers the value below the fit there and leaves certification
untouched.

\subsection*{The ring}

The ring of twelve spinless fermions at half filling was written on twelve qubits with the Jordan-Wigner
transformation, with the boundary condition for which the ground state is nondegenerate, which needs no
extra sign on the closing bond at fixed particle number. The Mermin value of the exact ground state was
maximized over the $24^n$ logical Pauli settings by coordinate ascent, agreeing with exhaustive enumeration at
$n=3$ and $4$. The hardware circuit starts from the domino superpositions of the cells and applies three
layers, each of controlled phases $e^{i\gamma n_in_j}$ on the bonds within and between dominoes followed by
hopping rotations $e^{-i\beta(XX+YY)/2}$ on the same bonds, with four angles per layer optimized for fidelity
with the exact ground state, in $114$ two qubit pulses. The first layer's phases within dominoes act on states
with one particle per domino, where they are the identity, and are omitted. The hopping rotation is
compiled as $R_{XX}(\beta)$ followed by $R_{XX}(\beta)$ conjugated by $S$. The runner's measured observables
equal $Z_{r_0}$, $X_{r_0}X_{r_1}$ and $Y_{r_0}X_{r_1}$ on the full four dimensional domino space, so the measured
quantity is the one computed for the ground state. The fixed settings, optimal for the strong coupling cat,
reproduce the optimum at every $V$. Each point used $32$ settings of $4096$ shots, and the
margin that assumes nothing about the noise from Hoeffding's inequality on the full sum of shots is $0.476$ at $\delta=10^{-6}$.
Before the run, stochastic trajectory simulation with two qubit depolarizing noise after every pulse
predicted visibilities of $0.44$, $0.30$ and $0.25$ at $0.4\%$, $0.7\%$ and $1.0\%$ error per pulse, so the
measured $0.41$ on \texttt{ibm\_fez} and $0.52$ to $0.54$ on \texttt{ibm\_kingston} correspond to about $0.45\%$
and $0.35\%$. Full tables of every run, the open chain and quench comparisons, and the preparation fidelities
at every layer count are in the Supplementary Information.

\subsection*{Open problems}

The first is the constant for arbitrary settings. Every numerical search over arbitrary measurement planes returns
$2\sqrt2$, while the certified bound is $3.26$ and the linear programming relaxation gives $3.12$. The
relaxation attains $3.12$ at an isolated corner of the angle box, where its first derivative vanishes and
its decrease is quadratic, which is why interval branch and bound converges only as a small power of run time
(Supplementary Information). Certifying $3.12$ needs a second order argument at that corner, and closing the remaining gap to $2\sqrt2$
needs the local parity constraint that the relaxation discards.

The second is the angle resolved law for $n>3$ and for commuting settings with disjoint planes. The Pfaffian
argument of Theorem~3 bounds the maximum but not the full angle dependence, and the helper mode argument that
does resolve it has so far been carried through for three parties.

The third is the thermodynamic limit of the transition witness. The onset $V^*(L)$ moves toward $V_c$ over
the four sizes accessible to exact diagonalization, and whether it reaches $V_c$, and how, is a question for
larger rings.

The fourth is quantitative. The robustness bounds above concern a resource measure, while classical simulation
cost is governed by the extent~\cite{ReardonSmith2024,DiasKoenig2024}. Connecting a measured violation to the
extent would turn the certificate into a lower bound on the classical cost of reproducing the data.

\section*{Data availability}
The raw shot records of every run reported here, including runs not quoted in the main text, are deposited
with the code below.

\section*{Code availability}
Code reproducing every result of this article, the protocol derivation, the exact reference simulator, the noise model, the light cone audit, the certified linear programming bound, the attaining constructions, the
ring model and its ground state circuit, the hardware runner for all three cells with a self test that
validates every compiled circuit against the exact reference, and the analysis scripts that recompute every
quoted number from the raw bitstrings, will be deposited in a public repository with a DOI upon publication,
and is available from the author in the meantime.

\section*{Acknowledgements}
The author acknowledges support from the Helmholtz Association HGF (Germany), the Hamburgische Investitions- und F\"orderbank (IFB) (Germany), and the Ministry of Science, Research and Culture of the State of Brandenburg within the Center for Quantum Technologies and Applications (CQTA) (Germany). The experiments were run on IBM Quantum systems accessed through DESY.

\section*{Competing interests}
The author declares no competing interests.

\newgeometry{margin=1in}
\onecolumn
\singlespacing
\begin{center}
{\LARGE\bfseries Supplementary Information}\\[0.6em]
{\large Genuine multipartite nonlocality is fermionic magic}\\[0.4em]
Jamal Slim, Deutsches Elektronen-Synchrotron DESY, 22603 Hamburg, Germany
\end{center}
\setcounter{section}{0}\renewcommand{\thesection}{S\arabic{section}}\renewcommand{\theHsection}{S\arabic{section}}
\renewcommand{\theHsubsection}{S\arabic{section}.\arabic{subsection}}
\setcounter{equation}{0}\renewcommand{\theequation}{S\arabic{equation}}\renewcommand{\theHequation}{S\arabic{equation}}
\setcounter{figure}{0}\renewcommand{\thefigure}{S\arabic{figure}}\renewcommand{\theHfigure}{S\arabic{figure}}
\setcounter{table}{0}\renewcommand{\thetable}{S\arabic{table}}\renewcommand{\theHtable}{S\arabic{table}}
\renewcommand{\theHtheorem}{S\arabic{theorem}}\renewcommand{\theHlemma}{S\arabic{lemma}}\renewcommand{\theHconjecture}{S\arabic{conjecture}}

\section{Notation}
Majorana operators $c_1,\dots,c_N$ satisfy $\{c_a,c_b\}=2\delta_{ab}$. A fermionic Gaussian state, pure or
mixed, is determined by $\Gam_{ab}=\tfrac{i}{2}\avg{[c_a,c_b]}$, real antisymmetric with
$\|\Gam\|_{\mathrm{op}}\le1$ ($\Gam^2=-1$ iff pure). Wick's theorem gives, for linear forms
$g_i=w_i\!\cdot\!c$,
\begin{equation}
\avg{g_1\cdots g_{2m}}=\Pf(G),\qquad G_{ij}=w_i\!\cdot\!w_j-i\,w_i^{\top}\Gam w_j\ \ (i<j).
\end{equation}
Party $j$ owns a block $V_j$ of $2k$ Majoranas. In class E2 each setting is a bilinear
$O=i(u\!\cdot\!c)(v\!\cdot\!c)$, $u\perp v$ unit in $V_j$. In class E1 each setting is a local Gaussian rotation, readout of all
$k$ occupation numbers, and an arbitrary Boolean function $f$ from $\{\pm1\}^k$ to $\{\pm1\}$. E1 $\supset$ E2.

\section{Structural lemmas}

\begin{lemma}[Pfaffian correlator]\label{lempf}
For bilinears $O_j=i(u_j\!\cdot\!c)(v_j\!\cdot\!c)$ on disjoint blocks with orthonormal $(u_j,v_j)$,
$\avg{O_1\cdots O_n}=\Pf(W\Gam W^{\top})$, where $W$ has rows $u_1,v_1,\dots,u_n,v_n$.
\end{lemma}
\begin{proof}
$w_i\!\cdot\!w_j=0$ for $i\neq j$, so $G=-i\,W\Gam W^{\top}$ and $\Pf(G)=(-i)^n\Pf(W\Gam W^{\top})$, and the
$i^n$ carried by the observables cancels it.
\end{proof}

\begin{lemma}[Reduction]\label{lemred}
For class E2 the value depends on $\Gam$ only through its compression to
$\bigoplus_j\mathrm{span}(u_j^0,v_j^0,u_j^1,v_j^1)$, a $4n\times4n$ antisymmetric contraction, i.e.\ the
covariance of a mixed Gaussian state on $2n$ modes. Hence the bound is independent of the number of modes
per party and it suffices to treat $k=2$ with mixed states.
\end{lemma}

\begin{lemma}[Multilinearity]\label{lemmult}
$\Pf(W\Gam W^{\top})$ is alternating multilinear in the rows of $W$, so it factors through the bivectors
$\beta_j=u_j\wedge v_j$ and defines
$\Phi_\Gam(\beta_1,\dots,\beta_n)=\langle\beta_1\wedge\cdots\wedge\beta_n,\Gam^{\wedge n}/n!\rangle$,
extended complex linearly, with
$\bigl\langle\prod_j(O_j^0+iO_j^1)\bigr\rangle=\Phi_\Gam(B_1,\dots,B_n)$, $B_j=\beta_j^0+i\beta_j^1$.
\end{lemma}

\section{The inequality for vacuum and full occupation}

\begin{theorem}[Vacuum full bound]\label{thmvf}
For every fermionic Gaussian state on $M$ modes and every occupation pattern $n\in\{0,1\}^M$ with complement
$\bar n$,
\begin{equation}
P(n)P(\bar n)\le 2^{-M}\ (M\ \text{even}),\qquad
P(n)P(\bar n)\le 2^{-M-1}\ (M\ \text{odd}),
\end{equation}
and both are attained, by $\lfloor M/2\rfloor$ BCS pairs $\tfrac{1}{\sqrt2}(\ket{00}+\ket{11})$, plus one
maximally mixed mode when $M$ is odd.
\end{theorem}
\begin{proof}
By the Gaussian overlap formula $P(n)=2^{-M}\sqrt{\det(1-\Gam\Gam_n)}$, with $\Gam_n$ the covariance
matrix of $\ket{n}$, and likewise $P(\bar n)=2^{-M}\sqrt{\det(1+\Gam\Gam_n)}$ since $\bar n$ has
covariance $-\Gam_n$. Writing $S=\Gam\Gam_n$,
\begin{equation}
P(n)P(\bar n)=4^{-M}\sqrt{\det(1-S^2)} .
\end{equation}
Since $\Gam_n^{-1}=-\Gam_n$, we have $\det(\lambda-S)=\det(\lambda\Gam_n+\Gam)=\Pf(\lambda\Gam_n+\Gam)^2$
because $\lambda\Gam_n+\Gam$ is antisymmetric. The characteristic polynomial of $S$ is therefore the square
of a real polynomial $p$ of degree $M$, so the spectrum of $S$ consists of $M$ doubly degenerate eigenvalues
$\nu_1,\dots,\nu_M$ (the roots of $p$), closed under complex conjugation, with $|\nu_j|\le\|S\|\le1$. Hence
$\det(1-S^2)=\prod_j(1-\nu_j^2)^2$ with $|1-\nu_j^2|\le1+|\nu_j|^2\le2$, while for a real root
$0\le1-\nu_j^2\le1$. For odd $M$ the real polynomial $p$ has at least one real root, which supplies the
extra factor $\tfrac12$.
\end{proof}

\noindent\emph{Verification.} The overlap formula and the double degeneracy were checked against exact
Jordan-Wigner algebra on $6$ qubits, and the $M=3$ bound $\sqrt{P(n)P(\bar n)}\le1/4$ is attained to $10^{-6}$ by direct
optimization over mixed states (for a pure Gaussian state on an odd number of modes the product vanishes
identically, because $n$ and $\bar n$ have opposite parity).

\section{An elementary inequality}
\begin{lemma}\label{lemC}
For $s\in[0,1]^3$ and $\epsilon\in\{\pm\}^3$ put
$V(\epsilon)=\prod_p\sqrt{1+\epsilon_ps_p}+\prod_p\sqrt{1-\epsilon_ps_p}$, so $V(\epsilon)=V(\bar\epsilon)$.
For any two noncomplementary $\epsilon,\epsilon'$, $V(\epsilon)+V(\epsilon')\le4$.
\end{lemma}
\begin{proof}
Let $a_p=\sqrt{1+s_p}$, $b_p=\sqrt{1-s_p}$, so $a_p^2+b_p^2=2$, and let $D\neq\emptyset,[3]$ be the set of
coordinates where $\epsilon,\epsilon'$ differ. Then $V(\epsilon)+V(\epsilon')=T_D\,T_{D^c}$ with
$T_S=\prod_{p\in S}a_p+\prod_{p\in S}b_p$. For $|S|=1$, $T_S=a+b\le2$, and for $|S|=2$,
$T_S=a_1a_2+b_1b_2\le\sqrt{(a_1^2+b_1^2)(a_2^2+b_2^2)}=2$ by Cauchy-Schwarz. Since $|D|\in\{1,2\}$ both
factors are $\le2$.
\end{proof}

\section{Proof of Theorem 1 of the main text}\label{secthm1proof}
\begin{theorem}[Theorem 1 of the main text]
For every fermionic Gaussian state on any number of modes and all local Majorana bilinear settings in which
each party's two measurement planes either share a direction or commute,
$|\avg{Q_AQ_BQ_C}|\le2\sqrt2$, $Q_p=O_p^0+iO_p^1$. The bound is attained, and it tightens to $2$ when all
three parties use anticommuting settings.
\end{theorem}
\noindent\textbf{Corollary.} $|\avg{M_3}|\le2\sqrt2$ and $|\avg{S_3}|\le4$, where $M_3=\mathrm{Re}\avg{Q_AQ_BQ_C}$
is the Mermin polynomial and $S_3=M_3+\mathrm{Im}\avg{Q_AQ_BQ_C}=\sqrt2\,\mathrm{Re}(e^{-i\pi/4}\avg{Q_AQ_BQ_C})$
the Svetlichny polynomial. The first is the quantum bound for biseparable states of the Mermin inequality (Collins
\emph{et al.}, PRL \textbf{88}, 170405, and Bancal \emph{et al.}, PRL \textbf{106}, 250404), so free fermions cannot
certify genuine tripartite entanglement independently of the device. The second is Svetlichny's hybrid local bound
(Svetlichny, PRD \textbf{35}, 3066, and Seevinck and Svetlichny, PRL \textbf{89}, 060401), so free fermions are never
genuinely tripartite nonlocal in Svetlichny's sense. Note that $M_3\le2\sqrt2$ alone would \emph{not} imply the
latter, because a hybrid model with a Popescu-Rohrlich box between two parties reaches $M_3=4$. Both statements also
hold for convex mixtures of Gaussian states, by linearity of $\avg{Q_AQ_BQ_C}$ in $\rho$.

\begin{proof}[Proof of Theorem 1]
Reduce to two modes per party (Lemma~\ref{lemred}). Write $Q_p=O_p^0+iO_p^1$.

\emph{Type (a), settings sharing a Majorana direction.} $O^0=i\gamma_1\gamma_2$,
$O^1=i\gamma_1(\cos\varphi\,\gamma_2+\sin\varphi\,\gamma_3)$. Then the operator identity
$Q_p^\dagger Q_p=2(1+s_p\tilde Z_p)$ holds with the \emph{helper bilinear} $\tilde Z_p=i\gamma_2\gamma_3$
and $s_p=|\sin\varphi_p|$, where $\varphi_p=\pi/2$ is the anticommuting (Pauli type) case.
\emph{Type (b), commuting settings on disjoint planes.} Take any bilinear $\tilde Z_p$ anticommuting with
both settings, so that $s_p=0$.

In each local parity sector, $Q_p=q_+^{(p)}S_p^{+}+q_-^{(p)}S_p^{-}$ with $S^\pm$ the raising/lowering
operators of that sector in the $\tilde Z_p$ eigenbasis, and $|q_\pm^{(p)}|^2=2(1\pm s_p)$. The joint
eigenbasis of the three $\tilde Z_p$ together with the three complementary bilinears $\tilde Z_pP_p$ is an
occupation basis $\ket{n}$ of six rotated modes, and
\begin{equation}
\Phi=\sum_{k,\epsilon}\Bigl(\prod_pq^{(p)}_{\epsilon_p}\Bigr)\bra{\bar n_{k\epsilon}}\rho\ket{n_{k\epsilon}},
\end{equation}
where $k$ labels parity sectors and $\epsilon$ the helper occupations. Positivity of $\rho$ gives
$|\bra{\bar n}\rho\ket{n}|\le\sqrt{P(n)P(\bar n)}$, and summing over $k$ with Cauchy-Schwarz,
\begin{equation}
|\Phi|\ \le\ \sum_{\epsilon}w_\epsilon\,r_\epsilon,\qquad
w_\epsilon=\prod_p\sqrt{2(1+\epsilon_ps_p)},\quad r_\epsilon=\sqrt{P_a(\epsilon)P_a(\bar\epsilon)},
\end{equation}
with $P_a$ the occupation distribution of the three helper modes, which is a Gaussian marginal.
Theorem~\ref{thmvf} with $M=3$ gives $r_\epsilon\le1/4$, and Cauchy-Schwarz gives
$\sum_{\text{4 pairs}}r\le1/2$. Maximizing the linear form over this polytope places $1/4$ on the two
largest pair weights $W(\epsilon)=w_\epsilon+w_{\bar\epsilon}=2\sqrt2\,V(\epsilon)$, so
\begin{equation}
|\Phi|\ \le\ \tfrac14\cdot2\sqrt2\,\bigl(V_{(1)}+V_{(2)}\bigr)\ \le\ 2\sqrt2
\end{equation}
by Lemma~\ref{lemC}. If every $s_p=1$ only $\epsilon=+\!+\!+$ carries weight ($w=8$) and
$|\Phi|\le8\cdot\tfrac14=2$. Attainment. For two parties every phase rotation $\mathrm{Re}(e^{i\theta}Q_AQ_B)$ is a two setting
correlation Bell expression with quantum maximum $2$ (it equals $\mathrm{CHSH}/\sqrt2$ at $\theta=-\pi/4$ and
$\avg{a_0b_0}-\avg{a_1b_1}$ at $\theta=0$), so $|\avg{Q_AQ_B}|\le2$ for all quantum states, with equality at
the CHSH Tsirelson point, which free fermions reach with bilinear settings~[Clarke \emph{et al.}, PRX
\textbf{6}, 021005]. A third party holding a product mode and measuring the same bilinear in both settings
(planes sharing a direction, $s_C=0$) has $\avg{Q_C}=1+i$, $|\avg{Q_C}|=\sqrt2$. The product configuration
gives $|\avg{Q_AQ_BQ_C}|=2\sqrt2$, so the bound is attained, by a state that is separable across $AB|C$. If $C$
instead uses anticommuting settings ($s_C=1$) on a product state, $|\avg{Q_C}|=|\avg{O_C^0}+i\avg{O_C^1}|\le1$
and the product configuration gives $2$, consistent with the tightened bound. The numerical optimum
(Sec.~\ref{secnum}) has the same geometry, a strongly paired $AB$ block and a weakly attached $C$.
\end{proof}

\section{General planes, and $n=4$}\label{secgeneral}
For arbitrary planes the two parity sectors carry different angles $\theta_L=\theta_1-\theta_2$,
$\theta_R=\theta_1+\theta_2$. Splitting $Q_p$ by sector and applying Theorem~\ref{thmvf} to \emph{every}
subset $T$ of the six rotated modes (via $\sum_{\epsilon|_T=t}r_\epsilon\le\sqrt{P_T(t)P_T(\bar t)}\le
2^{-\lceil|T|/2\rceil}$, itself Cauchy-Schwarz on the fibre followed by Theorem~\ref{thmvf} on the Gaussian
marginal) gives a linear program in the $32$ pair weights. The relaxation is rigorous at every fixed angle vector.
Its maximum over the six sector angles was located by Nelder and Mead from random starts (\texttt{lp\_check2.py})
and is found at a corner, with value $\tfrac34+\tfrac{1}{2\sqrt2}=1.1036$, hence numerically
$|\Phi|\le2\sqrt2\times1.1036=3.12$ for all E2 settings. Because that outer maximization is numerical, we
also certify the constant by interval branch and bound (\texttt{lp\_branch\_bound.py}). On any box of
angles every pair weight is bounded above by evaluating its increasing factors
$\sqrt{1+\sin\theta}=\cos\tfrac\theta2+\sin\tfrac\theta2$ at the top of the box and its decreasing factors
$\sqrt{1-\sin\theta}$ at the bottom, and since the LP variables are nonnegative this bounds the LP over the
box with no Lipschitz constant. Party permutation symmetry reduces the root by $3!$. Best first refinement, checkpointed and resumable
(\texttt{lp\_branch\_bound.py}), certifies
\[
|\Phi|\le3.26\qquad\text{for all bilinear settings (rigorous)},
\]
with the largest open box at $1.154$ in units of $2\sqrt2$ after $2.7\times10^{5}$ LPs ($17$ min), still
decreasing toward $1.1036$. The certified constant improves monotonically with run time, but slowly. The
gap to $1.1036$ closes as roughly $t^{-0.29}$ in run time $t$, because the maximum sits at a corner of the
six dimensional box and the boxes around it must shrink in every direction at once. About a day of CPU
time reaches $3.18$, while certifying $3.12$ itself would take far longer than is practical. Closing the
last gap therefore needs a local argument at the maximizing corner, showing that the optimal vertex of the
linear program stays optimal in a neighborhood of it and that the corner is a local maximum there, with
branch and bound covering the rest of the box, where the gap is bounded away from zero.

\noindent The structure of the maximizer explains the slow convergence. The linear program attains
$\tfrac34+\tfrac1{2\sqrt2}$ at exactly eight of the $64$ corners of the angle box, those of the form
$(a,1-a,b,1-b,c,1-c)$ with $a,b,c\in\{0,1\}$, in which each party's two sectors sit at opposite ends of their
range, and takes the value $1$ or less at the other $56$. The maximum is isolated. Along every edge leaving a
maximizing corner the value decreases, but its first derivative in the angles vanishes and the decrease is
quadratic, about $0.09\,\theta^2$ per coordinate. A box of width $w$ around the corner is therefore
overestimated by $O(w)$ while the true gap inside it is only $O(w^2)$, so the required second order argument
is a negative definite Hessian of the linear program on the inward cone together with a bound on its third
derivatives. Either value excludes genuine
tripartite pseudotelepathy for all bilinear settings, but neither excludes a Svetlichny violation
($3.12\cdot\sqrt2=4.42>4$). The sharp constant, and hence the Svetlichny corollary, is
proven only under the hypothesis of Theorem~1 of the main text. The relaxation discards one exact constraint, namely that for general
planes $Q_p$ flips both helper occupations of party $p$ simultaneously and therefore conserves the local
parity $\tilde Z_p\tilde Z_p'$. Adding this constraint is the natural route to closing the gap to $2\sqrt2$.

The same LP constrained on subsets, written for $n$ parties with $2^{n-1}$ pair variables and for settings
satisfying the hypothesis of Theorem~1 of the main text, returns $2^{n/2}$ at $n=3$ and $n=4$ but $5.954$ at
$n=5$, above the true value $5.657$. The relaxation therefore does not extend beyond four parties. The
law for $n$ parties is proved instead by a different route, Theorem~3 of the main text, which bounds the Pfaffian
$\Pf(W^{\!\top}\Gam W)$ directly and gives $2^{n/2}$ for every $n$ whenever each party's two settings share a
direction. The script \texttt{n\_party\_law.py} checks the Pfaffian identity against the generic Wick
expansion to $10^{-16}$ and against state vector expectation values to $6\times10^{-16}$, and confirms the bound
and its attainment at $n=3,4,5$.

\begin{conjecture}[Law for $n$ parties with arbitrary planes]
$\max_{\text{Gaussian}}\bigl|\avg{\prod_{j=1}^nQ_j}\bigr|=2^{n/2}$ for all $n$ and all local bilinear
settings, including those whose two planes share no direction. Proved for direction sharing settings by
Theorem~3 of the main text, and verified numerically for arbitrary planes at $n=3,4,5$ to $10^{-6}$.
\end{conjecture}
\noindent Attainment in both phases of the Mermin family was checked by direct maximization over pure Gaussian
states and settings, using the Pfaffian formula. Write $R_n=\sqrt2\,\mathrm{Re}(e^{-i\pi/4}\prod_jQ_j)$ for the
rotated polynomial. It reaches $2.8284$ at $n=2$, the Tsirelson value of CHSH, and $5.6569$ at $n=4$, and the
Mermin polynomial reaches $4.0000$ at $n=4$ and $5.6569$ at $n=5$, in each case the value $2^{n/2}$ or
$\sqrt2\cdot2^{n/2}$ of Theorem~3. The attaining states are products of the block of Theorem~1 for three parties, of
Tsirelson pairs at phase $\pi/4$, and of pairs at phase zero, chosen so that the phases add to the phase of the
polynomial.

\noindent The hybrid local bound $2^{n-1}$ of the Svetlichny polynomial $S_n=\mathrm{Re}\prod_jQ_j+\mathrm{Im}\prod_jQ_j$
used in the main text was checked by exhaustion at $n=3,4,5$. In a hybrid model the parties split into two groups
whose outputs may depend on all inputs within the group, so each correlator factorizes as $g(x_G)h(x_H)$ with
$g,h$ arbitrary $\pm1$ functions. Maximizing over every bipartition and every such pair gives $4$, $8$ and $16$,
equal to $2^{n-1}$, against the free fermion cap $\sqrt2\cdot2^{n/2}$ of $4$, $5.66$ and $8$.

\noindent For arbitrary planes $Q_j$ no longer factors into two linear forms and the Pfaffian identity fails,
which is why the general case remains open. At $n=3$ it is certified to $3.26$ and located numerically in the
relaxation at $3.12$, against the value $2\sqrt2$ that every direct search returns.

\noindent A clean route that is \emph{exact but not tight} is the following. Cauchy-Schwarz on the pair sum gives
$|\Phi|\le2^{n/2}\sqrt{F(s)F(-s)}$ with $F(s)=\mathbb{E}_P[\prod_p(1+\epsilon_ps_p)]=\Pf(J+D_s\Gam D_s)$ by
the Pfaffian minor expansion ($D_s=\mathrm{diag}(\sqrt{s_p})$ per mode). One has
$\max\Pf(J+A)\Pf(J-A)=2^{2\lfloor n/2\rfloor}$ over antisymmetric contractions $A$ (itself a sharp
restatement of Theorem~\ref{thmvf}), so this route is tight only near $s=0$.

\noindent The monogamy bound and the four measured multi cell runs are shown in Fig.~3 of the main text.

\section{Proof of Theorem 2 of the main text}\label{secthm2}

Bring each party's settings to the canonical form $O_p^0=ic_1c_2$ and
$O_p^1=ic_1(\cos\varphi_p\,c_2+\sin\varphi_p\,c_3)$, $\varphi_p\in[0,\pi/2]$, and write $s_p=\sin\varphi_p$.

\emph{Upper bound.} Section~\ref{secthm1proof} gives, before its final inequality,
$|\Phi|\le\tfrac{\sqrt2}{2}[V_{(1)}+V_{(2)}]$ with $V(\epsilon)=\prod_p\sqrt{1+\epsilon_ps_p}+\prod_p\sqrt{1-\epsilon_ps_p}$
and $V_{(1)}\ge V_{(2)}$ the two largest of the four values on $\{\pm1\}^3$ modulo a global sign. The sum of the
two largest is the maximum of $V(\epsilon)+V(\epsilon')$ over pairs of distinct classes. Because $\epsilon$ and
$-\epsilon$ are identified, any two distinct classes have representatives differing in exactly one position $p$,
and then
\[
V(\epsilon)+V(\epsilon')=\Bigl[\sqrt{1+s_p}+\sqrt{1-s_p}\Bigr]
\Bigl[\sqrt{(1+\epsilon_qs_q)(1+\epsilon_rs_r)}+\sqrt{(1-\epsilon_qs_q)(1-\epsilon_rs_r)}\Bigr].
\]
With $\sqrt{1\pm\sin\varphi}=\cos\tfrac{\varphi}{2}\pm\sin\tfrac{\varphi}{2}$ on $[0,\pi/2]$ the first bracket is
$2\cos\tfrac{\varphi_p}{2}$, and the second is $2\cos\tfrac{\varphi_q-\varphi_r}{2}$ when $\epsilon_q=\epsilon_r$
and $2\cos\tfrac{\varphi_q+\varphi_r}{2}$ otherwise, the former being the larger. Hence
$V_{(1)}+V_{(2)}=4\max_p\cos\tfrac{\varphi_p}{2}\cos\tfrac{\varphi_q-\varphi_r}{2}$, which is the bound $B(\varphi)$.
The script \texttt{angle\_tightness.py} confirms that this closed form equals the value obtained from the
linear program to $10^{-15}$ at $2000$ random settings.

\emph{One party.} For any Gaussian state $\avg{Q_p}=\Gam_{12}(1+i\cos\varphi)+i\Gam_{13}\sin\varphi$ with
$(\Gam_{12},\Gam_{13})$ a vector of length at most one, so $|\avg{Q_p}|^2=x^2+(x\cos\varphi+y\sin\varphi)^2$, a
quadratic form with eigenvalues $1\pm\cos\varphi$. Thus $|\avg{Q_p}|\le\sqrt{1+\cos\varphi}=\sqrt2\cos\tfrac{\varphi}{2}$,
attained by the pure state pairing $c_1$ with the bisector $\cos\tfrac{\varphi}{2}c_2+\sin\tfrac{\varphi}{2}c_3$ of
the two settings and $c_4$ with the orthogonal direction.

\emph{Two parties.} Write $Q_p=ic_1^p\ell_p$ with $\ell_p=(1+i\cos\varphi_p)c_2+i\sin\varphi_p\,c_3$. The pure
state pairing $c_1^q$ with $c_1^r$, $c_4^q$ with $c_4^r$, and the $(c_2,c_3)$ plane of $q$ with that of $r$
through $R=\mathrm{Rot}(\sigma)\,\mathrm{diag}(1,-1)$, $\sigma=\tfrac{\varphi_q+\varphi_r}{2}$, leaves one Wick
contraction, $\avg{Q_qQ_r}=-\ell_q^{\!\top}R\,\ell_r$, and a direct evaluation gives
$|\ell_q^{\!\top}R\,\ell_r|=2\cos\tfrac{\varphi_q-\varphi_r}{2}$, which is also the upper bound for two parties.

\emph{Three parties.} On the product of the state of two parties on $\{q,r\}$ and the state of one party on $p$ the
Pfaffian factorizes, $\Phi=\avg{Q_qQ_r}\avg{Q_p}$, so $|\Phi|=2\sqrt2\cos\tfrac{\varphi_p}{2}\cos\tfrac{\varphi_q-\varphi_r}{2}$,
and choosing $p$ to maximize it gives $B(\varphi)$. The script \texttt{attainment\_construction.py} builds this
covariance matrix, verifies $\Gam^2=-1$ and $|\Phi|=B(\varphi)$ to $10^{-15}$ at $505$ settings. \hfill$\square$

\emph{Extremal sets and obtuse angles.} Each factor of $B$ is at most one, so $B=2\sqrt2$ exactly when some
$\varphi_p=0$ and $\varphi_q=\varphi_r$. A search over a $61^3$ grid finds the minimum $B=2$ at exactly the four
settings $(\tfrac\pi2,\tfrac\pi2,\tfrac\pi2)$ and the three permutations of $(\tfrac\pi2,0,0)$. For obtuse angles,
relabeling the outcome of one setting conjugates that party's $Q_p$ and brings its angle into $[0,\pi/2]$. The
upper bound depends only on $|\sin\varphi_p|$ and is unchanged, and direct maximization attains it at every
obtuse and mixed setting tested. Independent maximization with the settings held fixed, over pure Gaussian
states on two modes per party, reproduces $B(\varphi)$ to five decimals at $32$ canonical settings, ten chosen
and $22$ random. The Mermin polynomial alone, one phase of the correlator, is strictly below $B$ at generic
settings, for instance $2.33$ against $2\sqrt2$ at $s=(0.6,0.6,0)$, and equals it only at the Tsirelson point
and its permutations.

\noindent\emph{Encoded qubits satisfy the hypothesis.} Under the Jordan-Wigner map of a dual rail, $c_1=X_{r_0}$,
$c_2=Y_{r_0}$, $c_3=Z_{r_0}X_{r_1}$, the logical Paulis are $Z_L=-ic_1c_2$, $X_L=-ic_2c_3$ and $Y_L=ic_1c_3$. A
logical observable $\bm n\cdot\bm\sigma_L$ is therefore $\tfrac{i}{2}\sum_{ab}A_{ab}c_ac_b$ with $A$ a real
antisymmetric $3\times3$ matrix, which has rank two and equals $uv^{\top}-vu^{\top}$ for orthonormal $u,v$ spanning
the plane orthogonal to $\bm n$. Two such planes lie in the same three dimensional space, so their intersection has
dimension at least $2+2-3=1$, and the two settings share a direction. The dihedral angle between the planes is the
angle between their normals, which are the Bloch vectors, so $\varphi_p$ is the Bloch sphere angle between the two
settings. The same holds for a Majorana tetron qubit, whose logical Paulis are $i\gamma_a\gamma_b$ with
$a,b\in\{1,2,3\}$ at fixed total parity. At ten random pairs of logical measurements per party that are not Pauli, direct
maximization over Gaussian states returns $B(\varphi)$ of Theorem~2 exactly, with $\varphi_p$ the Bloch angle, and
never exceeds $2\sqrt2$.

\noindent\emph{More modes than the proof needs.} The reduction lemma says two modes per party suffice. As a check
on it, maximization over pure states with three modes per party, $18$ Majoranas and $153$ parameters, returns
$B(\varphi)$ to five decimals at $\varphi=(0.3,0.9,1.3)$ and $(0,0.8,0.8)$ and never exceeds it.

\section{Monogamy}
\begin{theorem}[Chaining]
Let $m$ Mermin cells occupy pairwise disjoint sets of modes, let $\rho$ be one arbitrary global Gaussian
state, and let every party measure a bilinear fixed by its setting, the settings of each cell drawn
independently and uniformly from its four even weight patterns. Let $W_i$ be the event that cell $i$ wins.
If every cell satisfies the hypothesis of Theorem~1 of the main text, then
$\Pr[\bigcap_{i=1}^{m}W_i]\le\beta^m$ with $\beta=\tfrac12+\tfrac{2\sqrt2}{8}=\tfrac{2+\sqrt2}{4}$, and
for arbitrary bilinear settings $\Pr[\bigcap_{i=1}^{m}W_i]\le\beta'^{\,m}$ with $\beta'=\tfrac12+\tfrac{3.26}{8}=0.9075$ (certified) and numerically $\tfrac12+\tfrac{3.12}{8}=0.890$.
\end{theorem}
\begin{proof}
Fix the settings. Each observable is, after a local Gaussian rotation, the occupation number of one rotated
mode, and the $3m$ measured modes are orthogonal, so the observables commute and may be measured in any order.
Conditioning a Gaussian state on occupation number outcomes of a subset of modes leaves a Gaussian state on
the remainder, and conditioning on a win event $W_j$ (a union of outcome patterns) leaves a convex mixture of
Gaussian states, on which the bound holds by linearity. The settings of cell $i$ are independent of everything
conditioned on, so averaging over them gives $\Pr[W_i\mid W_1\cdots W_{i-1},s]=\tfrac12+\avg{M_3}_{\rm cond}/8\le\beta$
uniformly by Theorem 1 applied to the conditional state. Therefore
$\Pr[\bigcap_iW_i]=\prod_i\Pr[W_i\mid\cdots]\le\beta^m$, and averaging over settings preserves the bound.
\end{proof}

\noindent\emph{Numerical confirmation.} Two cells, six parties, $k=2$, one global \emph{mixed} Gaussian state
with unrestricted covariance between the cells ($276+12+48=336$ real parameters, namely $\mathfrak{so}(24)$, Williamson
eigenvalues, and setting frames), exact Pfaffian correlators including the sixteen cross terms of $12$ Majoranas.
The product of two optimal single cells lies in the search space and attains $\beta^2$ to $7\times10^{-10}$.
Direct maximization from random initializations converges monotonically to $\beta^2=0.728553$ (value
$0.72853$ after $\sim1.4\times10^5$ function evaluations, gap $2\times10^{-5}$ and closing, with precision
limited by finite difference gradients) and never exceeds it. The converged states carry covariance blocks between the cells of Frobenius norm $\approx1.4$, i.e.\ the optimum is degenerate, and correlations between cells
can be present but cannot raise the value. The optimizer's default evaluation budget terminates a
run with $336$ parameters after about $44$ iterations, well short of convergence, which is why every run was
continued in warm restarted rounds as described in Sec.~\ref{secnum}.

\section{From Gaussian states to matchgate circuits}
\begin{lemma}[Light cone factorization, covariance form]
Let $U$ be a constant depth circuit of nearest neighbour matchgates with Majorana rotation $R\in O(2n)$, and
let input bits be loaded as computational basis excitations at qubits $q_A,q_B,q_C$ with pairwise disjoint
forward light cones. Then
\begin{equation}
\Gam(s)=M_A(s_A)M_B(s_B)M_C(s_C)\ \Gam_0\ (M_AM_BM_C)^{\top},\qquad M_p(s_p)=R\,O_p^{s_p}R^{\top},
\end{equation}
with each $M_p$ real orthogonal and supported inside party $p$'s cone.
\end{lemma}
\begin{proof}
$\ket{x}$ is Gaussian with $\Gam_x=O_x\Gam_0O_x^{\top}$, where $O_x$ is block diagonal and acts on the input
qubit's two dimensional Majorana block as the reflection $\mathrm{diag}(1,-1)$. Conjugating by $R$ gives
$M=RO R^{\top}$, orthogonal, and supported in the light cone because $R$ has that support structure.
\end{proof}

\noindent\textbf{Remark (a correction worth stating).} The naive operator level version of this lemma (``$X_p$
is a single Majorana, so matchgate conjugation keeps it linear'') is \emph{false}, because under Jordan-Wigner
$c_{2p}=(\prod_{j<p}Z_j)X_p$, a Majorana times a string of $Z$ operators. Loading an input bit with an $X$ gate is not a
Gaussian operation. Only the covariance statement above survives, and it is what the bound consumes. The
reflection has determinant $-1$ (an odd element, i.e.\ a Majorana operator rather than a rotation), which is
harmless at covariance level.

\noindent\emph{Numerical audit.} On random brickwork matchgate circuits, comparing the covariance
construction with exact Jordan-Wigner state vectors over all $8$ input patterns, $R$ is orthogonal to
$5\times10^{-14}$, $\mathrm{cov}(U\ket{0})=R\Gam_0R^{\top}$ holds to $4\times10^{-14}$, $M_p$ is orthogonal to
$10^{-13}$, the factorization holds to $3\times10^{-13}$, and the light cone leak (deviation of $M_p$ from the
identity outside the cone) is below $5\times10^{-14}$, for six circuits at depths $1$ and $2$, all of which pass.

\noindent\textbf{Consequence.} Each party sees a fixed Gaussian state, a local orthogonal transformation inside its own cone that depends on the setting, and occupation readout. If each party outputs one designated qubit, its
Heisenberg picture observable $U^\dagger Z U$ is a Majorana bilinear supported in its cone (class E2) and
the chaining theorem applies with $\beta'$, or with $\beta$ when the induced settings satisfy the hypothesis of
Theorem~1 of the main text.
If a party reads several qubits and postprocesses (class E1), only the numerical evidence of
Sec.~\ref{secnum} applies. The winning circuit uses, besides matchgates, two kinds of controlled phase gate, the
$\mathrm{CP}(\pi)$ that creates the equivalent to GHZ state and, when the settings are supplied as quantum
inputs, a controlled-$S_L=\mathrm{CP}(\pm\pi/2)$ between the input qubit and a rail (with classical control
of the settings, as in the relational problem model of Bravyi, Gosset and K\"onig, only $\mathrm{CP}(\pi)$ is
needed). Both circuit families in the separation are geometrically local, constant depth and classically
simulable, and differ only in the phase $\varphi$. The separation is in correlation power, and geometric
locality cannot be the separating resource.

\section{The dual rail protocol}
Logical qubit $=$ two adjacent physical qubits $(r_0,r_1)$, code space $\{\ket{01},\ket{10}\}$, with
$Z_L=Z_{r_0}$, $X_L=X_{r_0}X_{r_1}$, $Y_L=Y_{r_0}X_{r_1}$, all Majorana bilinears, hence
parity preserving. Logical $R_z$, $e^{i\theta X_L}$, $e^{i\theta Y_L}$ (hence $H_L$, $S_L$) and $\ket{+}_L$
preparation are matchgates. The only non-Gaussian element is
$\mathrm{CZ}_L(\varphi)=\mathrm{CPhase}(\varphi)=\mathrm{fSim}(0,\varphi)$, a matchgate iff $\varphi\equiv0$.
Because $Z_L=Z_{r_0}=-Z_{r_1}$ on the code space, the controlled phase may act on \emph{either} rail, so a
cell of three parties is a bare physical line of six qubits with $R_{XX}$ within dominoes and $\mathrm{CPhase}$ between dominoes, with no swaps and three two qubit layers, and it embeds on any heavy hex processor. On hardware
with fractional two qubit gates,
$\mathrm{CPhase}(\varphi)=e^{i\varphi/4}R_z^{(a)}(\tfrac\varphi2)R_z^{(b)}(\tfrac\varphi2)R_{zz}(-\tfrac\varphi2)$,
one calibrated pulse, with $R_{zz}(-\theta)=X_aR_{zz}(\theta)X_a$ covering the sign.

The graph state on the three vertex path is equivalent under local Cliffords to GHZ$_3$. An exhaustive search over all $216$
per party Pauli setting pairs finds four optimal assignments, and the one with the fewest basis change gates is
$(Y,Z)$ for $A$, $(X,Y)$ for $B$, and $(Y,Z)$ for $C$, with outcome signs $(+,+),(+,+),(-,+)$, i.e.\ party $C$ reports $-Y_L$ for
its first setting. The measured polynomial is therefore
\begin{equation}
M=-\avg{Y_AX_BY_C}-\avg{Y_AY_BZ_C}-\avg{Z_AX_BZ_C}+\avg{Z_AY_BY_C}
\end{equation}
(with the standard Mermin sign pattern and unrelabeled outcomes the same settings give $M\equiv0$). The ideal
correlators are $\avg{Y_AX_BY_C}=\avg{Y_AY_BZ_C}=\avg{Z_AX_BZ_C}=-\tfrac12(1-\cos\varphi)$ and
$\avg{Z_AY_BY_C}=+\tfrac12(1-\cos\varphi)$, so the signal is exactly
\begin{equation}
M(\varphi)=2\,(1-\cos\varphi)
\end{equation}
(verified to $10^{-9}$), giving analytic thresholds $\varphi_{\mathrm{LHV}}=\pi/2$ and
$\varphi^{\ast}=2\arcsin 2^{-1/4}=1.997875$~rad $=114.47^{\circ}$. The latter is the certification threshold independent of the device (settings untrusted). If the settings are trusted to be the anticommuting logical
Paulis actually implemented, Theorem~1 of the main text gives the Gaussian bound $2$ and certification begins at $\varphi=\pi/2$.
Magic (non-Gaussianity of the state) is present for every $\varphi\neq0$.

\emph{The path does not generalize, the star does.} For $n\ge4$ the path graph state is a linear cluster
state, \emph{not} equivalent under local Cliffords to GHZ, and its Mermin value saturates at $4$ (verified, with path $=4.000$ at
$n=3,4,5$, and star $=4.000$ at $n=3$ and $16.000$ at $n=5$). A cell of $n$ parties requires a star or tree, hence
depth $O(\log n)$, not constant depth. The star cell of four parties is specified and measured in
Sec.~\ref{secstar}.

\emph{The star cell of four parties.} One center domino is coupled by controlled phases to three leaf dominoes.
On a heavy hex processor a degree three node has only two neighbors besides its own partner, so the center
couples two leaves through its rail $r_0$ and the third through $r_1$. A controlled phase on the $r_1$ rail
acts, on the code space, as the logical controlled phase conjugated by $X_L$ on the center, which for general
$\varphi$ is a controlled phase of the opposite sign together with a local Gaussian phase on the leaf, so the
state is GHZ type up to local Cliffords at $\varphi=\pi$ and the Pauli settings must be derived for the circuit
as built. Exhaustive search over the $24^4$ per party Pauli pairs and outcome signs, using the exact reference
simulator and the same operational readout as the hardware, gives the optimum $8.000000$ at $\varphi=\pi$ for
the settings $(X,Y)$, $(Y,Z)$, $(Y,Z)$, $(Z,Y)$ with both outcomes of the fourth party negated, and the ideal
signal with those settings is exactly
\[
M_4(\varphi)=2\,(1-\cos\varphi)^2 ,
\]
monotone on $[0,\pi]$ and verified to $10^{-15}$ at nine angles. It crosses the free fermion bound $4$ at
$2\arcsin2^{-1/4}=114.47^\circ$, the same angle at which the signal of three parties $2(1-\cos\varphi)$ crosses
$2\sqrt2$, because $(1-\cos\varphi)^2=2$ and $1-\cos\varphi=\sqrt2$ are the same condition. The runner selects
this cell with \texttt{--geometry star4}, and its layout search requires every coupled pair to be a device
edge, which was checked on a synthetic heavy hex graph with up to eight disjoint cells before any hardware
use.

\section{Measured and predicted performance, and postselection}

\subsection{Hardware run}
The protocol was first executed on \texttt{ibm\_fez} on 9 September 2026, then on \texttt{ibm\_marrakesh}
and \texttt{ibm\_berlin}, with the run with sixteen cells and the star cells of four parties on \texttt{ibm\_fez} and the interacting
ring on \texttt{ibm\_fez} and \texttt{ibm\_kingston} in October 2026. Mitigation on every
run was dynamical decoupling (XpXm) and measurement twirling. Gate twirling was \emph{disabled} because the IBM
runtime rejects it in combination with the fractional gate compilation of $\mathrm{CP}(\varphi)$ (runtime
error 1519). Both are permitted by the fair play rules below but not simultaneously, and the single native
$R_{zz}$ pulse was judged the more valuable of the two. No readout mitigation and no zero noise
extrapolation were applied anywhere. Transpiler optimization level $1$. The exact configuration of each run
is recorded in its results file.

\paragraph{Summary of all runs.} Every hardware run performed for this work, without exception, with its
file in \texttt{data/hardware/}. The rows marked ``first run'' and ``superseded'' are included for
completeness and are not quoted in the main text. Gate twirling is on only when fractional gates are off, since the runtime forbids
the combination. All values are raw, with no postselection.

\begin{center}\scriptsize
\setlength{\tabcolsep}{3.1pt}
\begin{tabular}{@{}llccccll@{}}
\toprule
device & processor & cells & shots & settings & frac. & result & role\\
\midrule
\texttt{ibm\_fez} & Heron r2 & 1 & 8192 & 9-point scan & yes & $M=3.305$, $A=0.832$ & dial (quoted)\\
\texttt{ibm\_fez} & Heron r2 & 1 & 8192 & $\varphi=\pi$ & yes & $M=3.243$ & repeat\\
\texttt{ibm\_fez} & Heron r2 & 1 & 4096 & 9-point scan & yes & $M=3.340$, $A=0.843$ & repeat\\
\texttt{ibm\_fez} & Heron r2 & 1 & 8192 & $\varphi=\pi$ & no & $M=3.087$ & compilation check\\
\texttt{ibm\_fez} & Heron r2 & 1 & 4096 & $\varphi=\pi$ & no & $M=3.108$ & first run\\
\texttt{ibm\_fez} & Heron r2 & 1 & 4096 & $\varphi=\pi$ & no & $M=3.121$ & first run, repeat\\
\texttt{ibm\_fez} & Heron r2 & 8 & 8192 & independent, 40 & yes & $\Pr=0.503\pm0.002$ & monogamy (quoted)\\
\texttt{ibm\_fez} & Heron r2 & 8 & 8192 & common & yes & cells agree to $0.03$ & superseded\\
\texttt{ibm\_marrakesh} & Heron r2 & 6 & 8192 & independent, 40 & yes & $\Pr=0.645\pm0.003$ & 2nd device (quoted)\\
\texttt{ibm\_marrakesh} & Heron r2 & 8 & 8192 & independent, 40 & yes & $\Pr=0.370\pm0.005$ & two weak cells\\
\texttt{ibm\_berlin} & Nighthawk r1 & 1 & 8192 & 9-point scan & yes & $M=3.538$, $A=0.869$ & 3rd device (quoted)\\
\texttt{ibm\_berlin} & Nighthawk r1 & 8 & 8192 & independent, 40 & yes & $\Pr=0.457\pm0.002$ & 3rd device (quoted)\\
\texttt{ibm\_fez} & Heron r2 & 4 star & 8192 & 9-point scan & yes & $M_4=7.02,\,6.91,\,6.86,\,6.07$ & four parties (quoted)\\
\texttt{ibm\_fez} & Heron r2 & 16 & 800 & independent, 400 & yes & $\Pr=0.174\pm0.001$ & 96 qubits (quoted)\\
\texttt{ibm\_fez} & Heron r2 & 2 rings & 4096 & $V=6,20$ & yes & $M_6=9.44,\,12.27$ & ring pilot\\
\texttt{ibm\_fez} & Heron r2 & 2 rings & 4096 & 10-point $V$ scan & yes & $M_6$ up to $12.68$ & ring (quoted)\\
\texttt{ibm\_kingston} & Heron & 2 rings & 4096 & $V=6,20$ & yes & $M_6=12.86,\,17.78$ & ring pilot\\
\texttt{ibm\_kingston} & Heron & 2 rings & 4096 & 10-point $V$ scan & yes & $M_6$ up to $16.67$ & ring (quoted)\\
\bottomrule
\end{tabular}
\end{center}

\paragraph{Single cell, dial scan.} One cell on the physical line $[95,94,93,92,91,98]$, $8192$ shots per
setting, nine values of $\varphi$.

\begin{center}
\begin{tabular}{cccc}
\toprule
$\varphi/\pi$ & $M_{\mathrm{raw}}$ & ideal $2(1-\cos\varphi)$ & $M_{\rm raw}-2\sqrt2$\\
\midrule
$0.000$ & $-0.019$ & $0.000$ & \\
$0.125$ & $0.153$ & $0.153$ & \\
$0.250$ & $0.500$ & $0.585$ & \\
$0.375$ & $1.036$ & $1.234$ & \\
$0.500$ & $1.684$ & $2.000$ & \\
$0.625$ & $2.343$ & $2.765$ & \\
$0.750$ & $2.855$ & $3.414$ & $+0.03$\\
$0.875$ & $3.162$ & $3.848$ & $+0.33$\\
$1.000$ & $\mathbf{3.305}$ & $4.000$ & $\mathbf{+0.48}$\\
\bottomrule
\end{tabular}
\end{center}

\noindent With $\sigma(M)=2/\sqrt{8192}=0.022$ at worst case variance, the value at $\varphi=\pi$ exceeds
$2\sqrt2$ by $0.48$, 22 times the shot noise, and certification first occurs at $\varphi=0.875\pi$.
\emph{Reproducibility.} Three independent jobs at $\varphi=\pi$ with the fractional gate compilation gave
$M_{\rm raw}=3.340$ ($4096$ shots), $3.243$ and $3.305$ ($8192$ shots), each certifying, the weakest by
$0.41$. Their between runs standard deviation, $0.049$, is about twice the shot noise and reflects
calibration drift between jobs. It is far from the $0.47$ margin to the bound and does not affect
certification. The single cell line $[95,94,93,92,91,98]$ contains none of the qubits flagged in the calibration of
that day (qubit $133$ bad, qubit $72$ isolated, the $102$ to $103$ coupler dead). The layout search reads the
calibrated two qubit errors and avoided $72$, $102$ and $103$ in every run, while qubit $133$ was included
in cell~4 of the runs with eight cells, which had the lowest acceptance of the eight ($0.81$) and still certified
($M_{\rm raw}=3.369$). A single parameter fit of
$M=A\,2(1-\cos\varphi)$ gives
\[
A=0.8318\pm0.0036,\qquad\text{rms residual }0.026,\qquad\text{max residual }0.044,
\]
so the residuals sit at the level of the shot noise and the measured curve is the predicted cosine rescaled
by one visibility independent of $\varphi$, with no detectable shape distortion. This is the signature of
incoherent noise and is a nontrivial systematic check, since coherent errors would deform the shape. The
apparent thresholds follow from $\cos\varphi=1-\mathrm{thr}/(2A)$ and shift from $0.500\pi$ to $0.565\pi$
(local) and from $0.636\pi$ to $0.747\pi$ (Gaussian). An earlier scan at $4096$ shots gave
$A=0.8429\pm0.0044$ and $M_{\rm raw}=3.340$ at $\varphi=\pi$, consistent at the $2\sigma$ level.

The flagged data of every single cell run at $\varphi=\pi$ are compared with the corrected threshold of
Eq.~\eqref{eqT} below, with $\alpha_s$ the acceptance in each of the four settings. The first six rows are
\texttt{ibm\_fez}.

\begin{center}
\begin{tabular}{lcccccc}
\hline
run & fractional & shots & $M_{\rm raw}$ & $M_{\rm flag}$ & $T$ & $M_{\rm flag}-T$\\
\hline
dial scan (quoted) & yes & 8192 & 3.305 & 3.849 & 3.825 & $+0.023$\\
repeat & yes & 8192 & 3.243 & 3.801 & 3.857 & $-0.056$\\
earlier scan & yes & 4096 & 3.340 & 3.851 & 3.810 & $+0.041$\\
compilation check & no & 8192 & 3.087 & 3.784 & 4.112 & $-0.328$\\
first run & no & 4096 & 3.108 & 3.794 & 4.194 & $-0.400$\\
first run, repeat & no & 4096 & 3.121 & 3.788 & 4.174 & $-0.386$\\
\texttt{ibm\_berlin} dial scan & yes & 8192 & 3.538 & 3.917 & 3.539 & $+0.377$\\
\hline
\end{tabular}
\end{center}

\noindent For the quoted run the acceptances were $\alpha_s=(0.872,0.880,0.879,0.870)$, giving $C=3.328$ and
$T=3.825$. The flagged value exceeds $T$ there by $0.023$, against a statistical error of
$\sigma_{\rm flag}=\sigma_{\rm raw}/\sqrt{\alpha_{\min}}=0.024$, so it certifies nothing, and in the repeat it
falls below. With fractional gates the flagged value lies within two standard errors of $T$ in every \texttt{ibm\_fez}
run, and without them it falls short by about $0.35$, so on \texttt{ibm\_fez} flagged data certify nothing. On
\texttt{ibm\_berlin} the acceptances are $\alpha_s=(0.911,0.905,0.909,0.904)$, which lowers the threshold to
$T=3.539$, and the flagged value $3.917$ clears it by $0.377$, sixteen standard errors, and by more than the
margin $0.262$ at $\delta=10^{-6}$, which assumes nothing about the noise, computed conservatively with $N\alpha_{\min}$ shots per
setting. Flagged data thus certify on one device and not on the other, depending on the acceptance, while the
raw value certifies on both and needs no correction, which is why only raw values are quoted. The
uncorrected comparison $M_{\rm flag}>2\sqrt2$ would have been passed by about $1.0$ on every run, and on
\texttt{ibm\_fez} it would have been wrong.

\paragraph{Compilation check.} Repeats at $\varphi=\pi$ with $8192$ shots gave $M_{\rm raw}=3.243$
with fractional gates and $3.087$ without (with gate twirling enabled, since it is then allowed), margins of
$0.41$ and $0.26$ over the bound. Certification holds under both decompositions, and the single pulse
compilation is better by $0.16$ in $M$, as expected from its shorter sequence.

\paragraph{Monogamy, eight cells.} Eight cells on $48$ qubits, with each cell's settings drawn
\emph{independently and uniformly} ($40$ random configurations, $8192$ shots each, $3.28\times10^{5}$ shots
in total). Independence of the settings is required by Theorem~4 of the main text. A run in which all cells share one setting
tuple is not a valid test, because the correct ceilings are then $\beta$ and $0.75$ rather than $\beta^{m}$
and $0.75^{m}$.

\begin{center}
\begin{tabular}{ccccccccc}
\toprule
cell & 0 & 1 & 2 & 3 & 4 & 5 & 6 & 7\\
\midrule
$M_{\mathrm{raw}}$ & $3.269$ & $3.357$ & $3.245$ & $3.342$ & $3.369$ & $3.516$ & $3.460$ & $3.189$\\
margin over $2\sqrt2$ & $0.44$ & $0.53$ & $0.42$ & $0.51$ & $0.54$ & $0.69$ & $0.63$ & $0.36$\\
\bottomrule
\end{tabular}
\end{center}

\noindent Per cell values are the Mermin polynomial with the four settings weighted equally, each
setting's correlator pooled over the configurations in which that cell received it (about ten of the $40$,
so $\approx8\times10^{4}$ shots per setting). The worst case shot noise of that pooled estimator is
$\sqrt{\sum_s1/N_s}\approx0.007$, so the margins in the table are fifty to a hundred times it. Every cell certifies on its own, with mean $3.344$ and the smallest margin $0.36$ against a pooled shot noise
of $0.007$, and every cell exceeds
the single cell ceiling $\beta=0.8536$ on each of its four settings separately (minimum entry $0.891$). The
joint statistic is
\[
\Pr[\text{all 8 cells win}]=0.5028\pm0.0024,\qquad
\beta^{8}=0.2817,\qquad 0.75^{8}=0.1001,
\]
a factor $1.78$ above the Gaussian bound and $5.02$ above the $\NC$ bound.

\emph{On the error bar.} The settings are fixed within each configuration and different setting patterns
have measurably different success rates on the hardware. The spread of the $40$ per configuration joint
fractions is $0.0148$, against $0.0055$ expected from shot noise alone. The correct unit of replication is
therefore the configuration, not the shot, and the quoted uncertainty is the standard error over the $40$
configurations, $0.0024$ (a bootstrap over configurations gives $0.0023$). On this error the excess over
$\beta^{8}$, $0.221$, is $94$ times the standard error. The pooled binomial error, $0.0009$, would make it
$253$ times and is wrong by a factor $2.7$. We note it only because it is the number a naive analysis produces,
and we do not quote either ratio as a significance, since at this size the margin is set by systematics and
not by statistics. The worst single
configuration out of $40$ gives $0.476$, which is $0.19$ above $\beta^{8}$ against its own shot noise of
$0.0055$, so the conclusion does not depend on averaging over configurations at all.

\emph{Reproducibility of the cells.} Two independent eight cell runs on \texttt{ibm\_fez} (one with
common settings, one with independent settings) agree cell by cell to rms $0.029$ in $M_{\rm raw}$ (max
$0.062$).

\emph{Second device.} On \texttt{ibm\_marrakesh} an eight cell run with independent settings gave six cells
at $3.24$ to $3.55$ and two cells, placed on lower quality regions (acceptances $0.61$ and $0.73$ against
$0.85$ to $0.89$), at $1.82$ and $2.47$, which fail to certify individually. The joint probability
$0.370\pm0.005$ still exceeds $\beta^{8}$ by $0.09$. A six cell run on the first six cells of the same
layout ranked by calibration (chosen after the run with eight cells exposed the two weak regions, which we state
explicitly) gave every cell certifying, $M_{\rm raw}$ from $3.279$ to $3.558$ (mean $3.438$), and
\[
\Pr[\text{all 6 cells win}]=0.6446\pm0.0026,\qquad \beta^{6}=0.3867,
\]
a margin of $0.26$, and the per cell values agree with the run with eight cells to rms $0.056$. The eight cell result is a useful illustration that the benchmark resolves
regional gate quality on a chip.

\emph{Third device, and a second architecture.} \texttt{ibm\_berlin} is a Nighthawk~r1 processor, a
different architecture from the Heron~r2 devices above, so the protocol is tested across two processor
families and not only across chips. On \texttt{ibm\_berlin} (13 September 2026, line $[0,10,11,12,2,3]$) the
nine point scan gives $M_{\rm raw}=3.538\pm0.022$ at $\varphi=\pi$, $0.71$ above $2\sqrt2$ and $0.28$
above the certified bound $3.26$ for arbitrary settings, the highest single cell value of the three devices, with
visibility $A=0.869\pm0.005$ and acceptances of $0.90$ to $0.92$ across the whole scan. The residuals of the
one parameter fit are $1.7$ times the shot noise ($\chi^2/\mathrm{dof}=3.1$) with a monotone trend. A
constant offset of $-0.04$ absorbs most of it, consistent with a small readout bias rather than a shape
distortion, and being negative it is conservative for certification. The eight cell run with independent
settings gives $\Pr[\text{all 8 win}]=0.4567\pm0.0019$ against $\beta^{8}=0.2817$, every
configuration certifying (worst $0.432$), cells independent to $1.1\%$. Cell~0 is the scan line and
reproduces it at $3.517$. Cell~6, on the poorest region (acceptance $0.79$ against $0.90$), gives $2.758$
and fails individually, the same tail pattern as on \texttt{ibm\_marrakesh}. Against the certified
$\beta'^{\,8}=0.462$ the eight cell value falls $0.005$ short, within three standard errors, while clearing the numerical $0.394$
by $0.06$.

\emph{Against the bound for arbitrary settings.} The threshold $2\sqrt2$ is a theorem under the hypothesis of
Theorem~1, which the compiled logical Paulis satisfy exactly, since any two of $X_L,Y_L,Z_L$ share a
Majorana direction. With no assumption on the settings beyond bilinearity the certified threshold is $3.26$
(Sec.~\ref{secgeneral}), giving $\beta'=0.9075$, $\beta'^{\,8}=0.462$ and $\beta'^{\,6}=0.561$. The numerically
established values are $3.12$, $0.890$, $0.394$ and $0.498$. Against the certified constants, the
\texttt{ibm\_berlin} single cell certifies by $0.28$, the \texttt{ibm\_fez} eight cell joint by $0.04$, the
\texttt{ibm\_marrakesh} six cell joint by $0.08$, and the best cells on all three devices by $0.26$ to $0.30$,
all many times the respective uncertainties, while the \texttt{ibm\_fez} single cell scan top sits $0.04$
above $3.26$, within two shot noise widths, and the \texttt{ibm\_berlin} eight cell joint $0.005$ below
$\beta'^{\,8}$. Against the numerical $3.12$ every
quoted value clears with margin. Certifying $3.12$ exactly is an open analytic step, described in
Sec.~\ref{secgeneral}, and needs no new data.

\emph{Independence of the cells.} The product of the eight uniformly averaged single cell win probabilities is
$0.5037$ against the measured joint $0.5028$, within half a standard error, and the mean single cell value
predicts $(\tfrac12+M_{\rm raw}/8)^{8}=0.5021$. The per cell values agree with an earlier common settings run
to within $0.02$ per cell. The $48$ physical qubits are distinct and the per cell acceptances range from
$0.81$ to $0.90$, which accounts for the spread in per cell $M_{\rm raw}$.

\emph{Scope of the claim.} The configurations were drawn pseudorandomly (seed $0$) and fixed at compile
time, which is what Theorem~4 of the main text requires for a nonadversarial device. This is a hardware benchmark of the
monogamy law, not a loophole free test independent of the device against an adversary with advance access to the
circuits, and no such claim is made.

\subsection{Sixteen cells on 96 qubits}\label{sec16}

Sixteen cells on $96$ disjoint qubits of \texttt{ibm\_fez} on 8 October 2026, with each cell's settings drawn
independently and uniformly over $400$ configurations of $800$ shots, $3.2\times10^{5}$ shots in total, the same
budget as the runs with eight cells but ten times as many configurations, chosen so that a statement that assumes nothing about the noise
becomes possible. The joint statistic is
\[
\Pr[\text{all 16 cells win}]=0.1743\pm0.0008,\qquad \beta^{16}=0.0794,\qquad 0.75^{16}=0.0100,
\]
a factor $2.20$ above the Gaussian bound, with the error the standard error over configurations (bootstrap
$0.0008$). The spread of the $400$ per configuration fractions is $0.0163$ against $0.0134$ from shot noise
alone. The worst configuration gives $0.134$ and the best $0.230$. The product of the sixteen uniformly averaged
single cell win probabilities is $0.1723$ against the measured joint $0.1743$, so the cells are independent.

\begin{center}
\begin{tabular}{lcccccccc}
\hline
cell & 0 & 1 & 2 & 3 & 4 & 5 & 6 & 7 \\
$M_{\rm raw}$ & 3.54 & 3.36 & 3.37 & 3.27 & 3.19 & 2.90 & 3.39 & 3.53 \\
acceptance & 0.91 & 0.86 & 0.89 & 0.86 & 0.86 & 0.82 & 0.86 & 0.86 \\
\hline
cell & 8 & 9 & 10 & 11 & 12 & 13 & 14 & 15 \\
$M_{\rm raw}$ & 2.83 & 3.40 & 3.10 & 2.86 & 3.22 & 3.33 & 2.68 & 2.80 \\
acceptance & 0.82 & 0.87 & 0.84 & 0.81 & 0.82 & 0.80 & 0.72 & 0.80 \\
\hline
\end{tabular}
\end{center}

\noindent Thirteen cells certify individually and three do not, the weakest at $2.675$ on the cell with the
lowest dual rail acceptance, $0.72$. The acceptance tracks the per cell Mermin value with correlation $+0.81$
across the sixteen cells, as on the runs with eight cells of \texttt{ibm\_berlin} and \texttt{ibm\_marrakesh}. The
joint result does not require every cell to certify. The monogamy bound $\beta^{16}$ holds for any Gaussian
strategy on all $96$ qubits, and the measured value exceeds it by $0.095$ while the per configuration spread is
$0.016$.

\noindent\emph{Distribution free certificate.} Hoeffding's inequality on the $400$ configuration fractions gives a
margin $\sqrt{\ln(2/\delta)/800}$, which is $0.097$ at $\delta=10^{-3}$ and $0.135$ at $\delta=10^{-6}$, and
neither clears $\beta^{16}$ from $0.174$. Hoeffding uses only that the fractions lie in $[0,1]$. The empirical
Bernstein inequality of Maurer and Pontil uses the measured sample variance $V$ as well, with margin
$\sqrt{2V\ln(2/\delta)/n}+7\ln(2/\delta)/(3(n-1))$. With $V=0.0163^2$ and $n=400$ this is $0.048$, $0.062$ and
$0.089$ at $\delta=10^{-3}$, $10^{-4}$ and $10^{-6}$, so the run certifies the monogamy law against any Gaussian
strategy with no assumption on the noise distribution at confidence $1-10^{-4}$ by $0.113$ against $0.079$, and
at $1-10^{-6}$ by $0.085$. The $400$ fractions are independent and identically distributed, each configuration
being an independent uniform draw of the settings measured in its own circuit, which is what both inequalities
require.

\noindent Against the constant for arbitrary settings, $\beta'^{\,16}=0.9075^{16}=0.212$, the run does not certify,
nor is it expected to at these per cell visibilities. The compiled settings satisfy the hypothesis of Theorem~1
exactly, so $\beta$ is the applicable bound. The physical qubits, in rail order per cell, are listed in the
results file.

\subsection{Four party star cells}\label{secstar}

Four star cells on $32$ disjoint qubits of \texttt{ibm\_fez} on 7 October 2026, nine dial angles, eight settings per angle, $8192$ shots per setting, with the
same mitigation as every other run and no readout mitigation or extrapolation. All values are raw, over all
shots. The shot noise at worst case variance is $8/\sqrt{8\times8192}=0.031$ per value.

\begin{center}
\begin{tabular}{cccccc}
\hline
$\varphi/\pi$ & ideal $2(1-\cos\varphi)^2$ & cell 0 & cell 1 & cell 2 & cell 3\\
\hline
0.000 & 0.000 & -0.062 & -0.040 & -0.004 & 0.017 \\
0.125 & 0.012 & 0.008 & 0.043 & 0.005 & 0.046 \\
0.250 & 0.172 & 0.132 & 0.157 & 0.110 & 0.077 \\
0.375 & 0.762 & 0.698 & 0.591 & 0.616 & 0.440 \\
0.500 & 2.000 & 1.696 & 1.489 & 1.617 & 1.281 \\
0.625 & 3.824 & 3.219 & 2.919 & 3.164 & 2.671 \\
0.750 & 5.828 & 4.977 & 4.612 & 4.867 & 4.190 \\
0.875 & 7.403 & 6.425 & 6.207 & 6.351 & 5.488 \\
1.000 & 8.000 & 7.022 & 6.905 & 6.860 & 6.073 \\
\hline
\end{tabular}
\end{center}

\noindent Every cell exceeds the bound $4$ from $\varphi=0.75\pi$ on. At $\varphi=\pi$ the eight signed
correlators of each cell lie within a band of width $0.06$ (cell 0 from $0.850$ to $0.907$, cell 3 from
$0.730$ to $0.782$), so the loss is a single visibility and not a distortion dependent on the setting. The
single visibility fits $M_4=A\,2(1-\cos\varphi)^2$ give $A=0.866\pm0.005$, $0.830\pm0.013$, $0.849\pm0.005$ and
$0.737\pm0.009$, with rms residuals $0.057$, $0.158$, $0.059$ and $0.116$. Cells 0 and 2 sit at twice the shot
noise. Cells 1 and 3 show a monotone residual pattern, negative in the middle of the dial and positive at the
top by up to $0.27$, so the measured curve is steeper than the ideal shape there. Taking the visibility from the
top point alone gives $M_4(\pi)/8=0.878$, $0.863$, $0.857$ and $0.759$. Neither reading affects certification,
which holds at every angle above the threshold. The crossings of the bound, by linear interpolation between dial
points, are $122.5^\circ$, $126.9^\circ$, $123.5^\circ$ and $132.2^\circ$, against $121.3^\circ$, $123.6^\circ$,
$122.3^\circ$ and $130.3^\circ$ predicted from each cell's fitted visibility through
$1-\cos\varphi=\sqrt{2/A}$. The dual rail flag passes $95\%$ to $97\%$ of shots on every domino except one leaf
of cell 3 (qubits $68$ and $67$), which passes $91\%$, and that cell is the weakest. The all cells acceptance
over sixteen dominoes is $0.50$, consistent with $0.957^{16}$, and plays no role in the raw values. The
physical qubits of the four cells, in rail order $[c_0,c_1,a,a',b,b',d,d']$, were $143$, $136$, $144$, $145$,
$142$, $141$, $123$, $124$, then $105$, $106$, $104$, $103$, $117$, $125$, $107$, $97$, then $3$, $16$, $2$,
$1$, $4$, $5$, $23$, $22$, and $69$, $70$, $78$, $89$, $68$, $67$, $71$, $58$.

\noindent This is a measurement of Theorem~3 of the main text at $n=4$. The bound equals the local bound there,
so the certified quantity is a plain violation of the Mermin inequality of four parties, by $2.1$ to $3.0$ against
a shot noise of $0.03$, with the device reaching $76\%$ to $88\%$ of the quantum maximum. The script
\texttt{make\_fig\_star.py} recomputes every number in this section from the raw bitstrings and draws
Fig.~4 of the main text.

\subsection{Predicted performance}
Density matrix simulation of the exact circuit with two qubit depolarizing $p_2$ after every two qubit gate,
single qubit depolarizing $p_1=p_2/10$ after every single qubit gate (including virtual $R_z$, which is
conservative), and symmetric readout error $p_{\mathrm{ro}}$ on every physical qubit. $M_{\rm raw}$ uses the
rail-$r_0$ bit of every party regardless of the flags, $M_{\rm flag}$ postselects on all three dual rail
flags, and $T$ is the flagged data threshold of Eq.~\eqref{eqT} evaluated with the per setting acceptances.

\begin{center}
\begin{tabular}{cccccc}
\toprule
$p_2$ & $p_{\mathrm{ro}}$ & $M_{\mathrm{raw}}$ & $M_{\mathrm{flag}}$ & $\alpha$ & $T$\\
\midrule
$0$ & $0$ & $4.000$ & $4.000$ & $1.000$ & $2.828$\\
$0.002$ & $0.010$ & $3.702$ & $3.973$ & $0.932$ & $3.330$\\
$0.003$ & $0.015$ & $3.560$ & $3.956$ & $0.900$ & $3.594$\\
$0.005$ & $0.020$ & $3.393$ & $3.926$ & $0.865$ & $3.908$\\
$0.010$ & $0.020$ & $3.253$ & $3.861$ & $0.844$ & $4.123$\\
$0.020$ & $0.030$ & $2.806$ & $3.714$ & $0.757$ & $5.081$\\
\bottomrule
\end{tabular}
\end{center}

\noindent Under this model the noisy curve is the ideal one rescaled by a factor independent of $\varphi$
$r=M_{\rm raw}/M_{\rm ideal}=0.890$ at the operating point $(p_2,p_{\rm ro})=(0.003,0.015)$, so the apparent raw
threshold is $\cos\varphi^\ast_{\rm noisy}=1-\sqrt2/r$, i.e.\ $0.636\pi\to0.70\pi$. (The flagged,
uncorrected curve crosses $2\sqrt2$ at $0.64\pi$ and the flagged certification corrected for $\alpha$ starts at
$0.80\pi$, and neither is the headline.) With $4096$ shots per setting and worst case variance $1$ per
correlator, $\sigma(M)=2/\sqrt{4096}=0.031$, the predicted margin over $2\sqrt2$ at the operating point is $0.73$.
Comparing with the measured run, the observed visibility $A=0.843$ against the predicted $0.890$ places
\texttt{ibm\_fez} near $p_2\approx5\times10^{-3}$ in this model, so the prediction at
$p_2=3\times10^{-3}$ was optimistic by about $6\%$ in $M$.

\noindent\textbf{Noise tolerance.} Readout error dominates, since every correlator is multiplied by
$(1-2p_{\rm ro})^3$, which accounts for $0.087$ of the $0.11$ loss at the operating point. Raw certification
survives to $p_2=3.0\%$ if the readout error is held at $1.5\%$ (margin $M_{\rm raw}-2\sqrt2=+0.002$ at
$p_2=3.0\%$), but only to $p_2=0.9\%$ if readout degrades in proportion ($p_{\rm ro}=5p_2$). The
flagged certification corrected for $\alpha$ survives to $p_2=1.0\%$ (fixed readout) and $0.4\%$ (proportional).

\noindent\textbf{Postselection.} The dual rail parity flags detect every single qubit $X$ or $Y$ fault
(verified on $6/6$ injected faults). Postselection is nevertheless \emph{not} free for this bound. Let
$\alpha_s$ be the acceptance in setting $s$ and split each raw correlator as $E_s=E_s^{\rm acc}+E_s^{\rm rej}$
with $|E_s^{\rm rej}|\le1-\alpha_s$. A Gaussian strategy obeys $\sum_st_sE_s\le2\sqrt2$ (Theorem~1 of the main text). Write
$y_s=t_sE_s^{\rm acc}/\alpha_s\in[-1,1]$ and $C=2\sqrt2+\sum_s(1-\alpha_s)$, so that $\sum_s\alpha_sy_s\le C$
and $M_{\rm flag}=\sum_sy_s$. Let $N$ be the set of settings with $y_s<0$, let $u=\sum_{s\in N}|y_s|$, and let
$\alpha_{\min}=\min_s\alpha_s$. Then $\alpha_{\min}\sum_{s\notin N}y_s\le\sum_{s\notin N}\alpha_sy_s\le C+u$ and
$\sum_{s\notin N}y_s\le4-|N|$, so $M_{\rm flag}\le\min\{(C+u)/\alpha_{\min}-u,\ 4-|N|-u\}$. The first bound
increases with $u$ and the second decreases, so their minimum never exceeds their value at the crossing,
$C+(1-\alpha_{\min})(4-|N|)$, when $N\neq\emptyset$, and it equals $C/\alpha_{\min}$ when $N=\emptyset$. Hence
\begin{equation}
M_{\mathrm{flag}}\ \le\ T=\max\Bigl\{\frac{C}{\alpha_{\min}},\ C+3(1-\alpha_{\min})\Bigr\},\qquad
C=2\sqrt2+\sum_s(1-\alpha_s),
\label{eqT}
\end{equation}
which equals $2\sqrt2$ only when every $\alpha_s=1$, and which coincides with the exact value of the
four variable linear program on a fine grid of acceptances. The second branch matters only when
$\alpha_{\min}>C/3\approx0.94$. Two simpler expressions are not bounds. The mean acceptance form is exceeded
by up to $0.05$ for $\alpha_s\in[0.5,1]$, and $C/\alpha_{\min}$ alone is exceeded by up to $10^{-3}$ when all
acceptances are close to one, because a negative $y_s$ on one setting buys room for the others. Certification from flagged data means
exceeding $T$. Since the raw value is a bilinear whether or not a shot is flagged, it is the figure, without caveats,
of merit and the one we quote. Readout mitigation and zero noise extrapolation rescale correlators, are not
available to the Gaussian strategy, and must not enter the comparison, whereas dynamical decoupling and
measurement twirling leaves correlators unbiased and was used. Gate twirling is also unbiased but could not be combined with the fractional gate compilation, and was disabled, as described in the hardware run above.

\section{The witness at the charge density wave transition}\label{seccdw}

\emph{Model.} Spinless fermions on a ring of $L=2n$ sites at half filling,
$H=-J\sum_i(c_i^\dagger c_{i+1}+\mathrm{h.c.})+V\sum_in_in_{i+1}$, with the boundary condition for which the
ground state is nondegenerate, periodic for odd particle number and antiperiodic for even, so that the
Jordan-Wigner boundary bond carries the sign $(-1)^{N-1}$. The model is the XXZ chain under Jordan-Wigner and
is a Luttinger liquid for $V<2J$ and a charge density wave for $V>2J$ in the thermodynamic limit. Parties
are the $n$ adjacent site pairs $(2j-1,2j)$, read as dual rail qubits, so every logical measurement is a
Majorana bilinear in three Majoranas of the pair and the settings share a direction (Sec.~\ref{secthm2}).
The ground state is obtained by sparse diagonalization in the half filling sector, of dimension $20$, $70$,
$252$ and $924$ for $L=6$, $8$, $10$ and $12$.

\emph{Mermin value.} $M_n$ is maximized over the $24^n$ choices of two logical Paulis and two outcome signs
per party by coordinate ascent from twelve random starts. For $n=3$ and $4$ the result agrees with exhaustive
enumeration to all digits, and the ascent reaches the same optimum from every start for all four sizes. The
exact Mermin values at $V=20J$ are $3.93$, $7.84$, $15.60$ and $31.05$, against the cat values $4$, $8$, $16$
and $32$, with cat overlaps $0.982$, $0.980$, $0.975$ and $0.970$. At $V=0$ the Gaussian ground state gives
$M_n=1.00$ at every size, well inside the bound.

\begin{center}
\begin{tabular}{ccccc}
\hline
$V/J$ & $L=6$, $n=3$ & $L=8$, $n=4$ & $L=10$, $n=5$ & $L=12$, $n=6$\\
free fermion bound & $2.83$ & $4$ & $5.66$ & $8$\\
\hline
 0.0 & 1.00 & 1.00 & 1.00 & 1.00 \\
 1.0 & 1.39 & 1.65 & 1.97 & 2.33 \\
 2.0 & 1.83 & 2.54 & 3.53 & 4.91 \\
 2.5 & 2.06 & 3.05 & 4.56 & 6.83 \\
 3.0 & 2.27 & 3.59 & 5.72 & 9.17 \\
 3.5 & 2.48 & 4.13 & 6.95 & 11.78 \\
 4.0 & 2.67 & 4.64 & 8.15 & 14.44 \\
 5.0 & 2.99 & 5.51 & 10.25 & 19.13 \\
 6.0 & 3.23 & 6.16 & 11.78 & 22.53 \\
 8.0 & 3.53 & 6.93 & 13.55 & 26.42 \\
12.0 & 3.79 & 7.53 & 14.90 & 29.43 \\
20.0 & 3.93 & 7.84 & 15.60 & 31.05 \\
\hline
onset $V^*$ & $4.47$ & $3.38$ & $2.97$ & $2.76$\\
visibility needed at $V=20J$ & $0.72$ & $0.51$ & $0.36$ & $0.26$\\
\hline
\end{tabular}
\end{center}

\noindent The onset $V^*(L)$, located by bisection to $0.01J$, decreases monotonically toward $V_c=2J$. The
visibility needed is the bound divided by the measured value at strong coupling, and it falls as $2^{1-n/2}$
because the cat's value grows as $2^{n-1}$ while Theorem~3's bound grows as $2^{n/2}$.

\emph{Open chains.} On an open chain the same model does not produce a cat. With open boundaries $N$
particles on $2N$ sites can avoid every adjacent pair in more than two configurations, for instance
$101001$ and $100101$ at six sites, and single hops connect them without an energy cost, so the
strong coupling ground state spreads over all of them. Its overlap with the two pattern cat is $0.3$ and
its Mermin value saturates at $1.85$ for three parties, below the local bound. The witness therefore
distinguishes the ring's cat from the chain's spread.

\emph{Dynamics.} A quench of the open chain from the free fermion Fock state $\ket{101010}$ under the same Hamiltonian, for $V$ from $0$ to $8J$ and times to $4/J$, never raises the of three parties
Mermin value above $1.74$. Interacting dynamics scatters correlations over many modes, while a Mermin
violation needs the opposite, coherence concentrated between two global patterns. This is why the ground
state, and not a quench, is the right object for the witness.

\emph{Preparation on hardware.} Two variational layers of the form $e^{-i\beta_k H_{\rm hop}}e^{-i\gamma_k H_V}$
applied to the free fermion ground state reproduce the eight site ground state with fidelity $0.999$, $0.988$,
$0.996$, $0.995$, $0.956$ and $0.988$ at $V=2$, $3$, $3.4$, $4$, $6$ and $8J$, with Mermin values $2.47$,
$3.14$, $3.95$, $4.59$, $5.59$ and $6.84$ against the exact $2.54$, $3.59$, $4.02$, $4.64$, $6.16$ and $6.93$. One
layer is insufficient above $V=4J$. The free fermion ground state itself requires a Gaussian circuit of
depth $O(L)$. Starting instead from the product of domino superpositions used in the cells, two layers reach fidelity only $0.39$ to $0.76$ at twelve sites, while three layers reach $0.84$ to $0.99$ across the whole range, and the three layer circuit is the one run on hardware below. The star cells of four parties of
Sec.~\ref{secstar} are that limit for the eight site ring. The script \texttt{cdw\_ring.py} reproduces every
number in this section and draws Fig.~5 of the main text.

\emph{The twelve site ring on hardware.} The ring was measured on two hexagonal twelve qubit loops of
\texttt{ibm\_fez} in parallel on 8 October 2026, ring 0 on qubits $123$, $136$, $143$, $144$, $145$, $146$, $147$,
$137$, $127$, $126$, $125$, $124$ and ring 1 on qubits $47$, $48$, $49$, $50$, $51$, $58$, $71$, $70$, $69$, $68$,
$67$, $57$, in ring order. Each circuit starts from the domino superpositions of the cells and applies three
layers, each consisting of controlled phases $e^{i\gamma n_in_j}$ on the bonds within and between dominoes and hopping
rotations $e^{-i\beta(XX+YY)/2}$ on the same bonds, with four angles per layer optimized for fidelity with the exact
ground state by \texttt{ring\_design.py}. The first layer's phases within dominoes act on states with one particle per
domino, where they are the identity, and are omitted, which leaves $114$ two qubit pulses. The hopping rotation is
compiled as $R_{XX}(\beta)$ followed by $R_{XX}(\beta)$ conjugated by $S$, exact to $3\times10^{-16}$, and every angle
lies in the native fractional range. The runner's operational observables equal $Z_{r_0}$, $X_{r_0}X_{r_1}$ and
$Y_{r_0}X_{r_1}$ on the full four dimensional domino space, not only on the code space, so the measured quantity is
the one computed above. The fixed settings are $(X,Y)$, $(Y,X)$, $(Y,X)$, $(X,Y)$, $(X,Y)$, $(X,Y)$ with outcome
signs $(+,+)$, $(-,+)$, $(+,-)$, $(-,-)$, $(-,-)$, $(+,+)$, and they reproduce the per point optimum of the exact
ground state at every $V$. Values below are raw, over all shots, $32$ settings of $4096$ shots per point, shot noise
$0.088$.

\begin{center}
\begin{tabular}{ccccc}
\hline
$V/J$ & prepared, ideal & ground state & ring 0 & ring 1\\
\hline
0 & 0.88 & 1.00 & 0.49 & 0.19 \\
1 & 2.62 & 2.33 & 1.11 & 0.52 \\
2 & 5.96 & 4.91 & 2.64 & 0.96 \\
3 & 10.51 & 9.17 & 4.35 & 1.88 \\
4 & 17.22 & 14.44 & 6.84 & 5.03 \\
5 & 21.05 & 19.13 & 8.40 & 6.05 \\
6 & 23.75 & 22.53 & 9.87 & 7.10 \\
8 & 26.75 & 26.42 & 11.30 & 6.51 \\
12 & 28.95 & 29.43 & 11.59 & 6.91 \\
20 & 31.28 & 31.05 & 12.68 & 4.32 \\
\hline
\end{tabular}
\end{center}

\noindent Ring 0 follows the ideal prepared curve times one visibility, $A=0.408$, with rms residual $0.19$, and
crosses the bound $8$ at $V=4.74J$ by linear interpolation, against $4.62J$ predicted from $A$ as the point where the
prepared value reaches $8/A$. Hoeffding's inequality applied to the full sum of the $32\times4096$ independent
outcomes gives a margin $\sqrt{2K\ln(2/\delta)/N}$ that assumes nothing about the noise, with $K=32$ settings and $N=4096$ shots, which
is $0.345$ at $\delta=10^{-3}$ and $0.476$ at $\delta=10^{-6}$. Ring 0 exceeds the bound by $0.40$, $1.87$, $3.30$,
$3.59$ and $4.68$ at $V=5$, $6$, $8$, $12$ and $20J$, so every point from $6J$ certifies at $1-10^{-6}$ and the point
at $5J$ at $1-10^{-3}$. The dual rail acceptance per domino at $V=20J$ is $0.92$, $0.94$, $0.92$, $0.92$, $0.80$ and
$0.89$ on ring 0, and $0.90$, $0.54$, $0.88$, $0.91$, $0.88$ and $0.80$ on ring 1, whose second domino, qubits $49$ and
$50$, is faulty. Ring 1 is not a rescaling of the prepared curve, with ratios from $0.14$ to $0.30$, and certifies at
no $V$. A pilot on the same loops one job earlier gave $9.44$ and $12.27$ on ring 0 at $V=6$ and $20J$, and $6.97$
and $4.22$ on ring 1, against $9.87$, $12.68$, $7.10$ and $4.32$ in the scan, so both rings are reproducible and the
fault of ring 1 is in the hardware. Before the run, stochastic trajectory simulation with two qubit depolarizing
noise after every pulse predicted $A=0.44$, $0.30$ and $0.25$ at errors of $0.4\%$, $0.7\%$ and $1.0\%$ per pulse.
The measured $0.41$ corresponds to about $0.45\%$. The script \texttt{make\_fig\_ring.py} recomputes every number
in this paragraph from the raw bitstrings and draws Fig.~6 of the main text.

\emph{A second processor.} The same circuit was run on two twelve qubit loops of \texttt{ibm\_kingston} on
8 October 2026, ring 0 on qubits $21$, $22$, $23$, $24$, $25$, $37$, $45$, $44$, $43$, $42$, $41$, $36$ and ring 1
on qubits $47$, $48$, $49$, $50$, $51$, $58$, $71$, $70$, $69$, $68$, $67$, $57$. The labels of ring 1 coincide with
those of the faulty \texttt{ibm\_fez} loop because both chips number their heavy hex lattice the same way, and
the qubits are different devices. Values are raw, $32$ settings of $4096$ shots, shot noise $0.088$.

\begin{center}
\begin{tabular}{cccc}
\hline
$V/J$ & prepared, ideal & ring 0 & ring 1\\
\hline
0 & 0.88 & 0.34 & 0.49 \\
1 & 2.62 & 1.25 & 1.33 \\
2 & 5.96 & 2.95 & 3.13 \\
3 & 10.51 & 5.61 & 5.48 \\
4 & 17.22 & 9.19 & 9.97 \\
5 & 21.05 & 11.00 & 12.05 \\
6 & 23.75 & 12.56 & 13.44 \\
8 & 26.75 & 13.35 & 14.96 \\
12 & 28.95 & 15.25 & 15.48 \\
20 & 31.28 & 16.67 & 15.56 \\
\hline
\end{tabular}
\end{center}

\noindent The single visibility fits give $A=0.523$ and $0.541$, with ratios of measured to ideal between $0.48$
and $0.53$ on ring 0 and between $0.50$ and $0.58$ on ring 1 for $V\ge J$. The rings cross the bound at $V=3.67J$
and $3.56J$, against $3.71J$ and $3.64J$ predicted from their visibilities, and exceed it at every point from $V=4J$
by more than the margin $0.476$ at $\delta=10^{-6}$ that assumes nothing about the noise, the largest excess being $8.67$. Every domino
passes its dual rail flag on $0.88$ to $0.95$ of shots at $V=20J$. A pilot one job earlier gave $12.86$ and $16.54$
on ring 0 at $V=6$ and $20J$, against $12.56$ and $16.67$ in the scan, and $13.76$ on ring 1 at $V=6J$ against
$13.44$. Ring 1 at $V=20J$ gave $17.78$ in the pilot and $15.56$ in the scan, a drop of $2.2$ in the last circuits of
the job with no change in any domino's flag rate, which we attribute to drift on that loop during the run. It does
not affect any conclusion, since both values exceed the bound by more than $7$. Across the three loops without a
fault the crossing moves from $4.74J$ at visibility $0.41$ to $3.6J$ at $0.53$, each time where the visibility
predicts, toward the ground state's $2.76J$.

\section{Numerical methods}\label{secnum}
All optimizations are L-BFGS-B with finite difference gradients and random restarts over the covariance
parametrization $\Gam=O\Gam_0O^{\top}$, $O=\exp(A)$ with $A$ antisymmetric, and $\Gam_0$ block diagonal with
entries $\tanh(\lambda_j)$ for mixed states. Correlators are exact Pfaffians (skew symmetric Gaussian
elimination, checked against a brute force perfect matching expansion and against $\Pf^2=\det$). Every
Wick/Pfaffian routine was validated against exact Jordan-Wigner linear algebra before use. Regarding termination, the
optimizer's default evaluation budget (\texttt{maxfun}$=15000$ in SciPy) terminates a finite difference run with $336$ parameters after $\approx44$ iterations, so all runs reported here were continued in warm restarted
rounds until the gain per round fell below $10^{-9}$ or the stated budget was exhausted, and the termination
message was logged. Norms of covariance blocks are Frobenius norms (every block of a contraction has operator
norm $\le1$). The principal results are collected in Table~S1.

\begin{table}[h]
\centering
\caption*{\textbf{TABLE S1.} Principal numerical results. Class E2 means one Majorana bilinear per party and
setting, class E1 means readout of all occupation numbers with Boolean postprocessing. Restarts are
independent random initializations of the continuous parameters. Entries marked $^{a}$ are local searches
initialized at the bilinear optimum (see footnote below the table).}
\vspace{4pt}
\small
\begin{tabular}{@{}>{\raggedright\arraybackslash}p{1.95in}>{\raggedright\arraybackslash}p{1.1in}>{\raggedright\arraybackslash}p{0.85in}>{\raggedright\arraybackslash}p{1.6in}@{}}
\toprule
Quantity maximized & Class & Restarts & Result\\
\midrule
$\max|\avg{M_3}|$, arbitrary planes, $k=2$, $3$, $4$ modes per party & E2, pure states & $40$, $24$, $12$ & $2.828427$ in every run\\[3pt]
$\max|\avg{M_3}|$, arbitrary planes & E2, mixed states & $8$ & $2.828427$\\[3pt]
$\max|\avg{M_3}|$, Boolean postprocessing by coordinate ascent & E1, $k=2$ & $10^{\,a}$ & $2.828427$\\[3pt]
$\max|\avg{M_3}|$, all $16^{6}$ Boolean functions enumerated & E1, $k=2$ & $1^{\,a}$ & $2.828427$\\[3pt]
$\max|\avg{M_4}|$ & E2, pure states & $12$ & $4.000000$\\[3pt]
$\max|\avg{M_5}|$ & E2, pure states & $6$ & $5.656854$\\[3pt]
$\max\Pr[\text{two cells win}]$, direct search & E2, one global mixed state & $1$, run to convergence & $0.72853$, which is $\beta^{2}-2\times10^{-5}$\\[3pt]
$\max\Pr[\text{two cells win}]$, product ansatz & E2 & none needed & $0.728553$, equal to $\beta^{2}$ to $10^{-9}$\\
\bottomrule
\end{tabular}
\end{table}
\noindent $^{a}$\,The E1 searches are local. The continuous part (state and local rotations) is initialized at the
bilinear optimum and the $\pm1$ postprocessing functions are then optimized, by coordinate ascent from
$12$ random function assignments per restart or by exact enumeration of all $16^6$ assignments at every
step of a continuous reoptimization of $60$ iterations, followed by reoptimization of the continuous part.
These runs show that no local deformation of the bilinear optimum benefits from multiqubit readout. They are
not a global search over class E1, which has $84$ continuous and $24$ Boolean parameters at $k=2$.

\noindent Code implementing every result of this article (protocol derivation, exact reference simulator,
noise model, the light cone audit, the branch and bound certificate of the general planes constant, the
star cell of four parties, the interacting ring and its ground state circuit, their figures, and a hardware
runner for all three cells with a self test that validates the
compiled circuits against the exact reference before submission), together with the raw shot records of every run listed above, will be deposited in
a public repository with a DOI upon publication, as stated in the main text.

\end{document}